\documentclass[aps,prd,twocolumn,showpacs,floatfix,preprintnumbers,amsmath,amssymb,nofootinbib, notitlepage]{revtex4-1}

\input epsf
\usepackage{graphicx}
\usepackage{color}
\usepackage{mathtools}
\usepackage{mathbbol}
\usepackage{longtable}
\usepackage{multirow}
\usepackage{empheq}

\newcommand{\beq}{\begin{equation}}
\newcommand{\eeq}{\end{equation}}
\newcommand{\barr}{\begin{eqnarray}}
\newcommand{\earr}{\end{eqnarray}}

\newcommand{\rme}{\textrm{e}}

\newcommand{\bs}{\boldsymbol}

\newcommand{\aap}{Astron. Astrophys.}

\newcommand{\mnras}{Mon. Not. R. Astron. Soc.}

\newcommand{\physrep}{Phys. Rep.}

\newcommand{\jcap}{J. Cosm. Astropart. Phys.}

\usepackage{color}
\newcommand{\newversion}[1]{{#1}}

\newcommand{\lsim}{\mathrel{\hbox{\rlap{\lower.55ex\hbox{$\sim$}} \kern-.3em \raise.4ex \hbox{$<$}}}}
\newcommand{\gsim}{\mathrel{\hbox{\rlap{\lower.55ex\hbox{$\sim$}} \kern-.3em \raise.4ex \hbox{$>$}}}}

\begin{document}
\title{Boltzmann-Fokker-Planck formalism for dark-matter--baryon scattering}
\author{Yacine Ali-Ha\"imoud}
\affiliation{Center for Cosmology and Particle Physics, Department of Physics,
New York University, New York, NY}
\date{\today}

\begin{abstract}

Linear-cosmology observables, such as the Cosmic Microwave Background (CMB), or the large-scale distribution of matter, have long been used as clean probes of dark matter (DM) interactions with baryons. It is standard to model the DM as an ideal fluid with a thermal Maxwell-Boltzmann (MB) velocity distribution, in order to compute the heat and momentum-exchange rates relevant to these probes. This approximation only applies in the limit where DM self-interactions are frequent enough to efficiently redistribute DM velocities. It does not accurately describe weakly self-interacting particles, whose velocity distribution unavoidably departs from MB once they decouple from baryons. This article lays out a new formalism required to accurately model DM-baryon scattering, even when DM self-interactions are negligible. The ideal fluid equations are replaced by the collisional Boltzmann equation for the DM phase-space distribution. The collision operator is approximated by a Fokker-Planck operator, constructed to recover the exact heat and momentum exchange rates, and allowing for an efficient numerical implementation. Numerical solutions to the background evolution are presented, which show that the MB approximation can over-estimate the heat-exchange rate by factors of $\sim 2-3$, especially for light DM particles. A Boltzmann-Fokker-Planck hierarchy for perturbations is derived. This new formalism allows to explore a wider range of DM models, and will be especially relevant for upcoming ultra-high-sensitivity CMB probes.  

\end{abstract}

\maketitle

\section{Introduction}

Possible non-gravitational interactions between dark matter (DM) and standard-model (SM) particles are testable with cosmological observations, complementing collider \cite{Boveia_18} direct-detection \cite{Undagoitia_16} and indirect \cite{Gaskins_16} searches. Linear-cosmology observables, in particular the cosmic microwave background (CMB), provide especially clean tests of such interactions, as the underlying physics is, in principle, very well understood, and computationally tractable. In addition to being sensitive to energy injection from annihilating or decaying DM \cite{Chen_04, Slatyer_09}, the CMB is a powerful probe of elastic scattering between DM and SM particles \cite{Dubovsky_01, Boehm_01}. On one hand, the resulting heat exchange between the DM and the photon-baryon fluids can lead to distortions of the CMB blackbody spectrum \cite{ACK15}. On the other hand, momentum exchange between the DM and any of the SM fluids (photons, neutrinos, or electron-baryons) affects the linear evolution of initial perturbations, and leaves an imprint on CMB temperature and polarization anisotropies. Heat and momentum exchange through DM-baryon scattering also affect the matter power spectrum and derived observables, such as the Lyman-$\alpha$ forest \cite{Dvorkin_14}; they also leave characteristic imprints on the high-redshift 21-cm brightness temperature and its fluctuations \cite{Tashiro_14, Munoz_15}.

The first quantitative study of the effect of DM scattering on CMB anisotropies and the matter power spectrum was carried out in \cite{Boehm_02}, specifically for energy-independent DM-photon interactions. The authors of \cite{Boehm_02} modified Euler's equations for the photons and the DM, assumed to be an ideal fluid, by adding a drag term between the two species, analogous to Compton drag between photons and the electron-baryon fluid. Shortly after, \cite{Chen_02} derived analogous equations for DM scattering off baryons with a velocity-independent cross section. In order to compute the heat and momentum-exchange rates, which depend on the velocity distributions of the interacting particles, they assumed that DM, like baryons, has a Maxwell-Boltzmann (hereafter, MB) distribution. These studies have since then been extended to a variety of interactions, such as millicharged-DM \cite{Dubovsky_04}, DM with an electric dipole moment \cite{Sigurdson_04}, and DM scattering with neutrinos \cite{Mangano_06}. The problem has received increased interest in the last few years, following the release of \emph{Planck}'s high-sensitivity and high-resolution CMB maps \cite{Planck_2013, Planck_2015, Planck_2018}. Using Markov-Chain Monte-Carlo (MCMC) analyses of \emph{Planck} data, several groups derived high-precision bounds to the elastic cross section of DM with SM particles \cite{Wilkinson_14a, Dvorkin_14, Wilkinson_14b, Gluscevic_17, Xu_18, Boddy_18, Slatyer_18, Boddy_18b}. Still, even these most recent studies rely on the simple approximation that the DM is an ideal thermal fluid. 

Even though it is the interactions with SM particles that dictate the characteristic rates of heat and momentum exchange, the precise values of these rates are functions of the detailed DM velocity distribution. The latter depends not only on interactions with SM particles, but also on self-interactions. Indeed, if the DM is efficiently self-interacting, its velocities get rapidly reshuffled to maximize entropy, leading to the MB distribution. On the other hand, if DM self-interactions are sufficiently weak, its velocity distribution unavoidably departs from MB as it thermally and kinematically decouples from baryons. This could be the case, for instance, if DM interactions are due to a small electric charge or dipole moment, or if the interacting particles make up a subdominant fraction of the total DM abundance. Therefore, existing linear-cosmology bounds to DM-SM scattering strictly apply only to strongly self-interacting DM. What's more, upcoming high-resolution CMB missions \cite{Abazajian_16} promise to be sensitive to cross sections about twenty times weaker than \emph{Planck} \cite{Li_18}. Should a positive detection of interacting DM finally be made, it will be extremely useful to glean even more information on DM properties, in particular on the strength of its self interactions. In this work we take the first steps in exploring this new dimension of DM interactions.


To go beyond the ideal-thermal-fluid approximation, one must solve the collisional Boltzmann equation for the DM distribution. In this paper, we study, for the first time, the collision operator for DM scattering off non-relativistic SM particles (in practice, thermal electrons or nuclei) with a velocity-dependent cross section. It is rather straightforward to write an exact integral collision operator, but the resulting integro-differential Boltzmann equation would be computationally demanding. We therefore derive a Fokker-Planck (hereafter FP) approximation to the DM-SM collision operator. While such an approximation has been amply studied in the context of kinematic decoupling of DM scattering off relativistic or light SM particles \cite{Bertschinger_06, Bringmann_07, Bringmann_09, Kasahara_09, Binder_16, Binder_17}, ours is the first work to derive the corresponding FP operator for scattering with non-relativistic baryons, for an arbitrary mass ratio. In that case, the fractional change in velocity per scattering event need not be small, and one must give up the standard approach of Taylor-expanding the collision integrand. Instead, we adopt a top-down method to construct the FP operator: starting from a general, number-conserving diffusion operator, we enforce that it satisfies detailed balance, and gives the correct momentum and heat exchange rates. We show that, when the DM distribution becomes narrow and the diffusion approximation fails, the exact momentum and heat-exchange rates become closed-form expressions of the DM bulk velocity and effective temperature, which our FP operator is constructed to recover. Our FP operator is therefore the best-possible diffusion approximation to the exact collision operator. It makes it possible to efficiently solve for the coupled evolution of DM and SM fluids, which would have been numerically challenging with an integral collision operator. The formalism developed here can therefore be incorporated into future MCMC analyses of precision linear-cosmology data, with a modest additional computational cost relative to the standard ideal-thermal-fluid approximation.

The rest of this paper is organized as follows. In Section \ref{sec:general}, we write down the general form of the collision operator for elastically-scattering DM, and discuss its most important properties. In Section \ref{sec:VT}, we study the DM velocity drift rate and diffusion tensor, and derive general expressions, as well as specific ones applying to power-law cross sections. In Section \ref{sec:mom-heat}, we focus on the momentum and heat-exchange rates, and study the regimes where they take on closed-form expressions. Section \ref{sec:FP} describes the top-down construction of the FP operator. In Section \ref{sec:background}, we study the evolution of the background distribution, which we evolve numerically to quantify its non-thermal distortions, and their impact on the heat-exchange rate. We lay out the Boltzmann-Fokker-Planck hierarchy for the evolution of perturbations in Section \ref{sec:pert}. After discussing limitations and extensions of our formalism in Section \ref{sec:late-coupling}, we conclude in Section \ref{sec:conclusion}. Appendix \ref{app:Helium} discusses the velocity dependence of DM scattering with Helium nuclei. In Appendix \ref{app:entropy}, we give a proof that the entropy is the unique functional that increase for any distribution, in the case of local self-interactions. To facilitate a quick read through this manuscript, we have framed the most important equations.

\section{General setup and definitions} \label{sec:general}

In this work we restrict ourselves to non-relativistic DM particles $\chi$ (which need not be all of the dark matter), with mass $m_\chi$, abundance $n_\chi$, and mass density $\rho_\chi = m_\chi n_\chi$, scattering off non-relativistic scatterers with corresponding properties $m_s, n_s, \rho_s$. \newversion{We denote the total mass by $M \equiv m_s + m_\chi$.} In practice, scatterers are either $\chi$ itself, or standard-model nuclei or electrons. We limit ourselves to elastic scattering, and in particular, do not consider number-changing interactions. In other words, we assume that the abundance of $\chi$ is already fixed, i.e.~that chemical decoupling occurs much before kinetic decoupling, of interest here. This standard assumption \cite{Gondolo_91}, holds for a variety of DM models \cite{Chen_01, Bringmann_09}, though not for all \cite{Binder_17}. We denote background quantities with an overline -- for instance, $\overline{n}_s \propto a^{-3}$ is the background density of $s$, where $a$ is the scale factor. Throughout the paper we use natural units $\hbar = c = 1$.

\subsection{Cross section}

The interaction is quantified by a differential cross-section $d \sigma_{\chi s}/d \Omega$. It only depends on the magnitude of the relative velocity $v_{\chi s} \equiv |\bs{v}_\chi - \bs{v}_s|$, and on the angle between $\hat{n} \equiv \bs{v}_{\chi s}/v_{\chi s}$ and $\hat{n}' \equiv \bs{v}_{\chi s}'/v_{\chi s}$, where primes denote quantities after the scattering event \newversion{(and recalling that $v_{\chi s}' = v_{\chi s}$)}. A particularly relevant quantity is the momentum-exchange cross section 
\beq
\overline{\sigma}(v_{\chi s}) \equiv \int d^2 \hat{n}' \frac{d \sigma_{\chi s}}{d \Omega} (1 - \hat{n} \cdot \hat{n}'). \label{eq:sigma-bar}
\eeq
In addition to providing general expressions, we shall also consider power-law dependencies of the form
\beq
\overline{\sigma}(v_{\chi s}) = \sigma_n ~ v_{\chi s}^n, \label{eq:sigma_n}
\eeq
where $n$ is an (even) integer. The index $n = -4$ corresponds to a Coulomb-like interaction, which would arise, for instance, if the DM has a small electric charge \cite{Goldberg_86}, up to logarithmic corrections. The case $n = -2$ corresponds to a DM with an electric dipole moment \cite{Sigurdson_04}. Even and positive power laws $n = 0, 2, 4, 6$ would arise from the non-relativistic operators considered in Refs.~\cite{Anand_13}, completing the set introduced in Ref.~\cite{Fitzpatrick_13}. While all of these operators lead to a power-law cross section with protons, the velocity dependence for interactions with helium nuclei is, in principle, more complex for spin-independent operators \cite{Gluscevic_17, Boddy_18}. However, we justify in Appendix \ref{app:Helium} that for the physical conditions relevant to cosmological studies, non-pointlike effects are negligible for Helium, and the DM-Helium cross section can also be approximated by a power law in relative velocity.

\subsection{Boltzmann equation and collision operator}

We denote by $f_\chi(\bs{v})$ the probability distribution of DM velocities, normalized such that
\beq
\int d^3 v~ f_\chi(\bs{v}) = 1.
\eeq
This distribution is $m_\chi^3/n_\chi$ times the DM phase-space density, and the Boltzmann equation for the latter can be rewritten as
\beq
\frac{d}{dt}(n_\chi  f_\chi) = n_\chi C[f_\chi], \label{eq:Boltzmann}
\eeq 
where $t$ is the proper time and $d/dt$ is the total derivative along a free-particle trajectory.

The collision operator is the sum over scatterers $s$ of individual collision operators $C_{\chi s}$. Each one is an integral operator of the form 
\barr
C_{\chi s}[f_\chi](\bs{v}_\chi) &=&  \int d^3 v_\chi' \Big{\{} f_\chi(\bs{v}_\chi') \Gamma_{\chi s}(\bs{v}_\chi' \rightarrow \bs{v}_\chi) \nonumber\\
&&~~~~~~~ - f_\chi(\bs{v}_\chi) \Gamma_{\chi s}(\bs{v}_\chi\rightarrow \bs{v}_\chi')  \Big{\}}, \label{eq:C_pchi}
\earr
where $\Gamma_{\chi s}(\bs{v}_\chi \rightarrow \bs{v}_\chi')$ is the differential scattering rate per final velocity volume element. It should be clear that this operator explicitly conserves the number of DM particles: 
\beq
\int d^3 v_\chi~ C_{\chi s}[f_\chi](\bs{v}_\chi) = 0. \label{eq:number}
\eeq
The differential scattering rate is explicitly given by
\barr
\Gamma_{\chi s}(\bs{v}_\chi \rightarrow \bs{v}_\chi') &=& n_s \int d^3 v_s ~d^3 v_s' ~f_s(\bs{v}_s)  \nonumber\\
&& ~~~ \times \mathcal{M}_{\chi s}(\bs{v}_\chi, \bs{v}_s; \bs{v}_\chi', \bs{v}_s'),~~~\label{eq:Gamma}
\earr
where $f_s$ is the velocity distribution of scatterers, and
\barr
&&\mathcal{M}_{\chi s}(\bs{v}_\chi, \bs{v}_s; \bs{v}_\chi', \bs{v}_s') \equiv M^2 m_s m_\chi \frac{d \sigma_{\chi s}}{d \Omega}  \nonumber\\
&&\times \delta_{\rm D} \left(\frac12 ( m_\chi v_\chi'^2 + m_s v_s'^2) - \frac12 (m_\chi v_\chi^2 + m_s v_s^2) \right)\nonumber\\
&& \times \delta_{\rm D} \left(m_\chi \bs{v}_\chi' + m_s \bs{v}_s' - m_\chi \bs{v}_\chi - m_s \bs{v}_s \right). \label{eq:M_chis}
\earr
Changing integration variables from $\bs{v}_s'$ to $\bs{v}_{\chi s}'$ and integrating over $v_{\chi s}'$, we may equivalently write it as
\barr
&&\Gamma_{\chi s}(\bs{v}_\chi \rightarrow \bs{v}_\chi')  \equiv  n_s \int d^3 v_s f_s(\bs{v}_s) v_{\chi s} \int d^2 \hat{n}'  \frac{d \sigma_{\chi s}}{d \Omega} \nonumber\\
&&~~~~~~~~~~\delta_{\rm D} \left(\bs{v}_\chi' - \bs{v}_\chi - \frac{m_s}{M} v_{\chi s}(\hat{n}' - \hat{n}) \right).\label{eq:Gamma2}
\earr
In this paper we will focus on the collision operator due to scattering with non-relativistic SM particles, specifically nuclei or electrons, which we refer to as ``baryons", following standard abuse of nomenclature. These particles efficiently scatter with themselves and as a consequence have a MB distribution. Moreover, they scatter efficiently with one another, hence have a common mean velocity $\bs{V}_b$ and temperature $T_b$ (but of course, different species of baryons have different mass $m_s$ and density $n_s$). Explicitly, the scatterers' velocity distribution is
\beq
f_s(\bs{v}) = \mathcal{G}(\bs{v} - \bs{V}_b; T_b/m_s), \label{eq:fs} \\
\eeq
where $\mathcal{G}$ is the Gaussian distribution,
\beq
\boxed{\mathcal{G}(\bs{w}; T/m) \equiv \left(\frac{m}{2 \pi T} \right)^{3/2} \exp\left(-\frac{m w^2}{2T} \right)}. \label{eq:MB}
\eeq
As a consequence, it is easy to show from Eq.~\eqref{eq:Gamma} that the differential rates $\Gamma_{\chi s}$ satisfy the following \emph{detailed balance} property:
\barr
\exp\left( - \frac{m_\chi (\bs{v} - \bs{V}_b)^2}{2 T_b}\right) \Gamma_{\chi s}(\bs{v} \rightarrow \bs{v}') \nonumber\\
= \exp\left( - \frac{m_\chi (\bs{v}' - \bs{V}_b)^2}{2 T_b}\right) \Gamma_{\chi s}(\bs{v}' \rightarrow \bs{v}).
\earr
This implies that the collision operator conserves the MB distribution at temperature $T_b$ and mean velocity $\bs{V}_b$ for the DM (with mass $m_\chi$):
\beq
C_{\chi s}\left[\mathcal{G}(\bs{v} - \bs{V}_b; T_b/m_\chi) \right] = 0. \label{eq:C-db}
\eeq

\section{Velocity drift vector and diffusion tensor} \label{sec:VT}

In this section we discuss important intermediate quantities: the rates of velocity drift and diffusion. In subsequent sections, we will relate them to the momentum and heat-exchange rates, and to the coefficients of the Fokker-Planck operator.

\subsection{Definitions and general expressions}

The velocity drift rate is defined as 
\beq
\frac{d \langle \Delta \bs{v}\rangle }{dt}\Big{|}_{\bs{v}} \equiv  \int d^3 v'~ \Gamma_{\chi s}(\bs{v} \rightarrow \bs{v}') (\bs{v}' - \bs{v}).\label{eq:drift-def}
\eeq
Similarly, the velocity diffusion tensor is  
\beq
\frac{d \langle \Delta v^i \Delta v^j \rangle}{dt}\Big{|}_{\bs{v}} \equiv \int d^3 v' \Gamma_{\chi s}(\bs{v} \rightarrow \bs{v}') (v' - v)^i (v' - v)^j. \label{eq:diff-def}
\eeq
Using Eq.~\eqref{eq:Gamma2}, we obtain
\barr
\frac{d \langle \Delta \bs{v}\rangle }{dt}\Big{|}_{\bs{v}} &=& \frac{\rho_s}{M} \int d^3 v_s f_s(\bs{v}_s) v_{\chi s}^2  \nonumber\\
&&~~~~~\int d^2 \hat{n}'  \frac{d \sigma_{\chi s}}{d \Omega} (\hat{n}' - \hat{n}).
\earr
The innermost integral is just $- \overline{\sigma}(v_{\chi s})\hat{n}$, where $\overline{\sigma}$ is the momentum-exchange cross section defined in Eq.~\eqref{eq:sigma-bar}. We therefore find
\barr
\frac{d \langle \Delta \bs{v}\rangle }{dt}\Big{|}_{\bs{v}} = - \frac{\rho_s}{M} \int d^3 v_s f_s(\bs{v}_s) v_{\chi s} \bs{v}_{\chi s} \overline{\sigma}(v_{\chi s}). ~~\label{eq:drift2}
\earr
Changing integration variables to $\bs{u} \equiv \bs{v}_{\chi s}$, and using Eq.~\eqref{eq:fs} for the MB distribution $f_s$, we arrive at
\barr
\frac{d \langle \Delta \bs{v}\rangle }{dt}\Big{|}_{\bs{v}} &=& - \frac{\rho_s}{M} \int d^3 u ~u ~ \bs{u}~ \overline{\sigma}(u)\mathcal{G}(\bs{w} - \bs{u}; T_b/m_s) , ~~~~\label{eq:drift-final}
\earr
where from here on we use the notation
\beq
\boxed{\bs{w} \equiv \bs{v} - \bs{V}_b}.
\eeq
It should be clear from Eq.~\eqref{eq:drift-final} that the velocity drift is parallel to $\bs{w}$, and moreover vanishes for $w = 0$. It therefore takes the following form,
\beq
\frac{d \langle \Delta \bs{v}\rangle }{dt}\Big{|}_{\bs{v}}  = - \frac{\rho_s}{M} \mathcal{A}(w; T_b/m_s) \bs{w}, \label{eq:drift-sym}
\eeq
where, explicitly, 
\begin{empheq}[box=\fbox]{align}
\mathcal{A}(w; T/m) &\equiv \int d^3 u ~u ~ \frac{\bs{u} \cdot \bs{w}}{w^2}~ \overline{\sigma}(u) \mathcal{G}(\bs{w} - \bs{u}; T/m), \label{eq:A-def}
\end{empheq}
where we have purposefully left $T/m$ without labels to keep this expression general. 

Similarly, we rewrite the velocity diffusion tensor as
\barr
\frac{d \langle \Delta v^i \Delta v^j \rangle}{dt}\Big{|}_{\bs{v}} = \left(\frac{m_s}{M}\right)^2 n_s \int d^3 v_s f_s(\bs{v}_s) v_{\chi s}^3 \nonumber\\
\int d^2 \hat{n}'  \frac{d \sigma_{\chi s}}{d \Omega} (\hat{n}' - \hat{n})^i (\hat{n}' - \hat{n})^j.
\earr
In general, this tensor depends on the $\ell \leq 2$ multipole moments of the differential cross section. Its trace, however, only depends on the momentum-exchange cross section:
\barr
\frac{d \langle (\Delta \bs{v})^2 \rangle}{dt}\Big{|}_{\bs{v}} = 2 \left(\frac{m_s}{M}\right)^2 n_s \int d^3 v_s f_s(\bs{v}_s) v_{\chi s}^3 \overline{\sigma}(v_{\chi s}).~~ \label{eq:D2}
\earr
Here again, this only depend on $w = |\bs{v} - \bs{V}_b|$: explicitly, 
\barr
\frac{d \langle (\Delta \bs{v})^2 \rangle}{dt}\Big{|}_{\bs{v}} = 2 \frac{m_s}{M} \frac{\rho_s}{M} \mathcal{B}(w; T_b/m_s), \label{eq:diff-sym}
\earr
where 
\beq
\boxed{\mathcal{B}(w; T/m)  \equiv  \int d^3 u ~u^3 \overline{\sigma}(u) \mathcal{G}(\bs{w} - \bs{u}; T/m)}. \label{eq:B-def}
\eeq

\subsection{Fluctuation-dissipation relation}

Although individually $\mathcal{A}$ and $\mathcal{B}$ depend on the specific cross section, one can derive a general relation between the two. We start by rewriting the Gaussian as
\beq
\mathcal{G}(\bs{w} - \bs{u}; T/m) = \rme^{- m w^2/2 T} \mathcal{G}(u; T/m) \rme^{m \bs{u} \cdot \bs{w}/T},
\eeq
so that we may rewrite $\mathcal{A}(w)$ and $\mathcal{B}(w)$ as follows:
\barr
\mathcal{B}(w; T/m) &=& \rme^{- m w^2/2 T} \int du ~u \mathcal{F}(u)~ \mathcal{I}_0\left(\frac{m w u}{T} \right), ~~~~\label{eq:D(w)}\\
\mathcal{A}(w; T/m) &=& \frac1{w}\rme^{- mw^2/2 T} \int du ~\mathcal{F}(u)~ \mathcal{I}_1\left(\frac{m w u}{T} \right), ~~~~\label{eq:gamma(w)}\\
\mathcal{F}(u) &\equiv& 4 \pi u^4 \overline{\sigma}(u) \mathcal{G}(u; T/m), \label{eq:calF}\\
\mathcal{I}_0(X) &\equiv& \frac12 \int_{-1}^1 d \mu ~\rme^{X \mu} = \frac{\sinh X}{X}, \\
\mathcal{I}_1(X) &\equiv& \frac12 \int_{-1}^1 d \mu ~\mu ~\rme^{X \mu} \nonumber\\
&=& \frac{(X -1)\rme^{X} - (X +1) \rme^{-X}}{2 X^2}.
\earr
Using $2 \mathcal{I}_1+ X \mathcal{I}_1' = X \mathcal{I}_0$, we find
\barr
\boxed{\mathcal{B}(w) = \frac{T}{m} \Big{(}3 \mathcal{A}(w)  + w \mathcal{A}'(w) \Big{)} + w^2 \mathcal{A}(w) }. \label{eq:fluct-diss}
\earr
This can be thought of as a generalized fluctuation-dissipation relation (see, e.g.~\cite{Thorne_Blandford}). In particular, at $w \rightarrow 0$, we find
\beq
\mathcal{B}(0; T/m) = \frac{3 T}{m} \mathcal{A}(0; T/m). \label{eq:fluct-diss-0}
\eeq

\subsection{Asymptotic limits}

Provided $\overline{\sigma}$ is scale-free (as is the case for a power-law cross section, for instance), then it is easy to see that $\mathcal{A}(w; T/m)$ and $\mathcal{B}(w; T/m)$ have a characteristic scale $w_* \sim \sqrt{T/m}$. For $w\gg \sqrt{T/m}$, we have
\barr
\mathcal{A}(w\gg \sqrt{T/m}) &\approx& w ~\overline{\sigma}(w), \label{eq:A-asympt}\\
\mathcal{B}(w \gg \sqrt{T/m}) &\approx& w^3~ \overline{\sigma}(w).\label{eq:B-asympt}
\earr

\subsection{Explicit expressions for power-law cross sections}

If the cross section is a power-law of the form \eqref{eq:sigma_n}, the coefficients $\mathcal{A}$ and $\mathcal{B}$ can be expressed in terms of the confluent hypergeometric function of the first kind:
\barr
\mathcal{A}(w; T/m) &=& c_n~\sigma_n \left(\frac{T}{m}\right)^{\frac{n+1}{2}}  \alpha_n(\sqrt{m/T}~w),\label{eq:A-pow-law}\\
\mathcal{B}(w; T/m) &=& 3 c_n~\sigma_n \left(\frac{T}{m}\right)^{\frac{n+3}{2}} \beta_n(\sqrt{m/T}~w), \label{eq:B-pow-law}\\
\alpha_n(x) &\equiv & ~ _1F_1\left(- \frac{n + 1}{2}, \frac52, - \frac{x^2}2  \right), \\
\beta_n(x)  &\equiv& ~ _1F_1\left(- \frac{n + 3}{2}, \frac32, - \frac{x^2}2 \right), \\
c_n &\equiv& \frac{ 2^{\frac{5 + n}{2}}}{3 \sqrt{\pi}} \Gamma(3 + n/2).\label{eq:cn}
\earr
Note that the same hypergeometric functions appear in the momentum and heating rate if the DM has a MB distribution \cite{Boddy_18b} -- see also Ref.~\cite{Munoz_15} for the case $n = -4$. The coefficients $\mathcal{A}$ and $\mathcal{B}$ are more fundamental quantities, however, independent of the DM distribution: here the relevant temperature and mass are those of the scatterers. We will show in the next section how they imply the expressions of Refs.~\cite{Munoz_15, Boddy_18b}.

\section{Momentum- and heat-exchange rates} \label{sec:mom-heat}

While the phase-space density $f_\chi$ contains the full information about the DM velocity distribution, it is useful to write equations for its first few moments. In particular, the rates of momentum and heat exchange, related to the first two moments, are most relevant for cosmological observables. 

\subsection{General expressions}

We define the DM bulk velocity $\bs{V}_\chi$ and \emph{effective} temperature $T_\chi$ such that
\barr
\bs{V}_\chi &\equiv& \int d^3 v ~ \bs{v} ~f_\chi(\bs{v}),\\
T_\chi &\equiv&\frac13 m_\chi \int d^3 v ~ \left(\bs{v} - \bs{V}_\chi \right)^2 ~f_\chi(\bs{v}). \label{eq:Tchi}
\earr
The momentum exchange rate (per unit volume) is 
\barr
\dot{\boldsymbol{\mathcal{P}}}_\chi &\equiv& \rho_\chi \dot{\bs{V}}_\chi\big{|}_{\chi s} = - \rho_b \dot{\bs{V}}_b \big{|}_{\chi s} \equiv - \dot{\boldsymbol{\mathcal{P}}}_b\nonumber \\
&=& \rho_\chi \int d^3 v ~ \bs{v} ~C_{\chi s}[f_\chi](\bs{v}) \label{eq:momentum1} \\
&=&  \rho_\chi  \int d^3 v ~f_\chi(\bs{v}) \frac{d \langle \Delta \bs{v} \rangle}{dt}\Big{|}_{\bs{v}}, \label{eq:momentum2a}
\earr
where, to get the third line, we inserted the collision operator \eqref{eq:C_pchi} into Eq.~\eqref{eq:momentum1}, used the symmetries of the integrand, and the definition \eqref{eq:drift-def} of the velocity drift vector. Note that it is the \emph{total baryon density} $\rho_b$ that multiplies $\dot{\bs{V}}_b$, even if only a specific species scatters with the DM. Indeed, baryons quickly share momentum among all species through frequent interactions. Substituting Eq.~\eqref{eq:drift-sym}, we obtain the following expression for the momentum-exchange rate:
\barr
\boxed{\dot{\boldsymbol{\mathcal{P}}}_\chi = -  \frac{\rho_s \rho_\chi}{M} \int d^3 v f_\chi(\bs{v}) \mathcal{A}(w; T_b/m_s) \left(\bs{v} - \bs{V}_b\right)}.\label{eq:momentum2}
\earr
Similarly, the DM heating rate per unit volume is 
\barr
\dot{\mathcal{Q}}_\chi &\equiv& \frac32 n_\chi \dot{T}_{\chi} \big{|}_{\chi s}  \nonumber\\
&=& \frac12 \rho_\chi \int d^3 v ~ (\bs{v} - \bs{V}_\chi)^2 ~C_{\chi s}[f_\chi](\bs{v}). \label{eq:heat1}
\earr
Using the symmetries of the integrand, this can be re-expressed in terms of the trace of the velocity diffusion tensor \eqref{eq:diff-def} and of the drift rate \eqref{eq:drift-def} as follows:
\barr
\dot{\mathcal{Q}}_\chi &=& \frac12 \rho_\chi \int d^3 v ~f_\chi(\bs{v}) \nonumber\\
&&\times \left(\frac{d \langle (\Delta \bs{v})^2 \rangle}{dt}\Big{|}_{\bs{v}} + 2 (\bs{v} - \bs{V}_\chi) \cdot \frac{d \langle \Delta \bs{v} \rangle}{dt}\Big{|}_{\bs{v}}\right).\label{eq:heat2a}
\earr
Substituting Eqs.~\eqref{eq:drift-sym} and \eqref{eq:diff-sym}, we arrive at
\begin{empheq}[box=\fbox]{align}
\dot{\mathcal{Q}}_\chi &=  \frac{\rho_s\rho_\chi}{M} \int d^3 v ~f_\chi(\bs{v}) \times \Big{(} \frac{m_s}{M} \mathcal{B}(w; T_b/m_s) \nonumber\\
& ~~~~~~~~~ - (\bs{v} - \bs{V}_\chi)\cdot (\bs{v} - \bs{V}_b) \mathcal{A}(w; T_b/m_s) \Big{)}. \label{eq:heat2}
\end{empheq}
We may similarly define the baryon heating rate $\dot{\mathcal{Q}}_b \equiv \frac32 n_b \dot{T}_b |_{\chi s}$. Conservation of total energy (arising from both thermal and bulk motions) implies that
\beq
\boxed{\dot{\mathcal{Q}}_\chi + \dot{\mathcal{Q}}_b + \bs{V}_{\chi b} \cdot \dot{\boldsymbol{\mathcal{P}}}_\chi  = 0}, \label{eq:energy-cons}
\eeq
where we used $\dot{\bs{\mathcal{P}}}_b = - \dot{\bs{\mathcal{P}}}_\chi$. Here again, it is the total baryon number density $n_b$ that appears in the baryon heat exchange rate, for baryons quickly share heat even if only a specific species scatters with the DM.

\subsection{Case of a MB distribution} \label{sec:rates-MB}

If the DM is thermalized by frequent self-interactions, its distribution is MB:
\beq
f_\chi(\bs{v}_\chi) = \mathcal{G}(\bs{v}_\chi - \bs{V}_\chi; T_\chi/m_\chi).
\eeq
In that case, the momentum and heat-exchange rates can be re-expressed in terms of the drift and diffusion rates evaluated at $T/m = T_b/m_s + T_\chi/m_\chi$, as we show now. 

We start by inserting Eq.~\eqref{eq:drift2} into Eq.~\eqref{eq:momentum1}, to get
\barr
\dot{\boldsymbol{\mathcal{P}}}_\chi = - \frac{\rho_s \rho_\chi}{M} \iint d^3 v_\chi d^3 v_s f_\chi(\bs{v}_\chi) f_s(\bs{v}_s) \nonumber\\
\times  v_{\chi s} \bs{v}_{\chi s} \overline{\sigma}(v_{\chi s}).~~
\earr
We rewrite the joint Gaussian distribution of $\bs{v}_\chi, \bs{v}_s$ as the joint Gaussian distribution of the independent variables $\bs{v}_{\chi s} \equiv \bs{v}_\chi - \bs{v}_s$ and 
\beq
\bs{v}_+ \equiv \sqrt{\frac{T_b m_\chi}{T_\chi m_s}} (\bs{v}_\chi - \bs{V}_\chi) + \sqrt{\frac{T_\chi m_s}{T_b m_\chi}} (\bs{v}_s - \bs{V}_b), 
\eeq
with means $\bs{V}_{\chi b}$ and zero, respectively, and both with variance 
\beq
v_{\rm th}^2 \equiv T_\chi/m_\chi + T_b/m_s. \label{eq:vth}
\eeq
Changing integration variables to $\bs{v}_+$ and $\bs{u} \equiv \bs{v}_{\chi s}$ and integrating over the former, we arrive at
\barr
\dot{\bs{\mathcal{P}}}_\chi &=& - \frac{\rho_\chi \rho_s}{M} \int d^3 u ~u~ \bs{u}~\overline{\sigma}(u)  \mathcal{G}(\bs{u} - \bs{V}_{\chi b}; v_{\rm th}^2).
\earr
Comparing with Eqs.~\eqref{eq:drift-final} and \eqref{eq:A-def}, we see that the last integral is $\mathcal{A}(w; v_{\rm th}^2) \bs{w}$, evaluated at $\bs{w} = \bs{V}_{\chi b}$, and substituting $T/m \rightarrow v_{\rm th}^2$:
\beq
\dot{\bs{\mathcal{P}}}_\chi = \frac{\rho_\chi \rho_s}{M}  \mathcal{A}(V_{\chi b}; v_{\rm th}^2)\times \left(\bs{V}_b - \bs{V}_\chi \right). \label{eq:momentum-MB}
\eeq
We proceed similarly for the heating rate, which we rewrite, using Eqs.~\eqref{eq:drift2} and \eqref{eq:D2} as
\barr
\dot{\mathcal{Q}}_\chi = \frac{\rho_s \rho_\chi}{M} \iint d^3 v_\chi d^3 v_s f_\chi(\bs{v}_\chi) f_s(\bs{v}_s)  v_{\chi s}  \overline{\sigma}(v_{\chi s})\nonumber\\
\times \left\{\frac{m_s}{M}  v_{\chi s}^2  - \bs{v}_{\chi s} \cdot (\bs{v}_\chi - \bs{V}_\chi)\right\}.~~~ \label{eq:heat-MB-interm}
\earr
We rewrite  
\beq
\bs{v}_\chi - \bs{V}_\chi = \frac{\sqrt{\frac{T_\chi T_b}{m_\chi m_s}}}{v_{\rm th}^2} \bs{v}_+ + \frac{T_\chi/m_\chi}{v_{\rm th}^2}\left(\bs{v}_{\chi s} - \bs{V}_{\chi b} \right),
\eeq
so that the last term in Eq.~\eqref{eq:heat-MB-interm} is
\barr
\frac{m_s}{M}  v_{\chi s}^2  - \bs{v}_{\chi s} \cdot (\bs{v}_\chi - \bs{V}_\chi) = \nonumber\\
 \frac{T_b - T_\chi}{M v_{\rm th}^2} v_{\chi s}^2 + \frac{T_\chi/m_\chi}{v_{\rm th}^2} \bs{v}_{\chi s} \cdot \bs{V}_{\chi b} - \frac{\sqrt{\frac{T_\chi T_b}{m_\chi m_s}}}{v_{\rm th}^2} \bs{v}_{\chi s} \cdot \bs{v}_+. ~~~~
\earr
Again, we change variables to $\bs{v}_+$, $\bs{u} \equiv \bs{v}_{\chi s}$, and integrate over $\bs{v}_+$. Here again, we can re-express the result in terms of the drift and diffusion rates evaluated at $w = V_{\chi b}$, substituting $T_b/m_s \rightarrow v_{\rm th}^2$:
\barr
\dot{\mathcal{Q}}_\chi &=&    \frac{\rho_\chi \rho_s}{M^2 v_{\rm th}^2} \mathcal{B}(V_{\chi b};v_{\rm th}^2) \times (T_b - T_\chi) \nonumber\\
&+&   \frac{\rho_\chi \rho_s}{M v_{\rm th}^2} \frac{T_\chi}{m_\chi} \mathcal{A}(V_{\chi b};v_{\rm th}^2) V_{\chi b}^2 . \label{eq:heat-MB}
\earr
Using Eq.~\eqref{eq:energy-cons}, we see that $\dot{\mathcal{Q}}_b$ takes the same expression, with exchange of $\chi$ and $s$, as it should. The first term in Eq.~\eqref{eq:heat-MB} corresponds to heat transfer from the hottest to the coolest component. The second term is a net heating for both components, due to dissipation of bulk motions into heat, with efficiency proportional to the thermal velocity dispersion of each fluid.

If we specialize to power-law cross sections, inserting Eqs.~\eqref{eq:A-pow-law}-\eqref{eq:B-pow-law} into Eqs.~\eqref{eq:momentum-MB} and \eqref{eq:heat-MB}, we recover the expressions derived in Ref.~\cite{Boddy_18b} (and Ref.~\cite{Munoz_15} for $n = -4$). Once again, Eqs.~\eqref{eq:A-pow-law}-\eqref{eq:B-pow-law} are more fundamental quantities, wich do not depend on the DM distribution.

\subsection{Limiting cases with closed forms} \label{sec:closed-form}

In general, the integrals appearing in the momentum and heat-exchange rates \eqref{eq:momentum2} and \eqref{eq:heat2} cannot be expressed as closed forms of $\bs{V}_\chi$ and $T_\chi$. They depend on the full phase-space distribution $f_\chi(\bs{v})$, which must be obtained by solving the full Boltzmann equation \eqref{eq:Boltzmann}. Besides the case where $f_\chi$ is a MB distribution, discussed above, closed forms may also be obtained if $f_\chi$ is sufficiently narrow. Throughout we assume a scale-free cross section, implying that $w^2 \mathcal{A}'' \sim w\mathcal{A}' \sim \mathcal{A}$ and similarly for $\mathcal{B}$. 

\subsubsection{Case where $T_\chi/m_\chi \ll V_{\chi b}^2$}

If $f_\chi(\bs{v}) = f_\chi(\bs{w} + \bs{V}_b)$ is sufficiently narrow relative to its mean at $w = V_{\chi b}$, we may Taylor-expand $\mathcal{A}$ and $\mathcal{B}$ around $w = V_{\chi b}$, and obtain, up to corrections of relative order $(T_\chi/m_\chi)/V_{\chi b}^2 $: 
\barr
\dot{\boldsymbol{\mathcal{P}}}_\chi &\approx& \frac{\rho_s \rho_\chi}{M} \mathcal{A}(V_{\chi b}; T_b/m_s) \times \left(\bs{V}_b - \bs{V}_\chi\right),\label{eq:momentum-narrow}\\
\dot{\mathcal{Q}}_\chi &\approx& \frac{\rho_s \rho_\chi}{M} \left(\frac{m_s}{M}\mathcal{B} - \frac{T_\chi}{m_\chi} \left(3 \mathcal{A} + w \mathcal{A}'\right)\right)_{(V_{\chi b}; T_b/m_s)}.~~ 
\earr
Using the fluctuation-dissipation relation \eqref{eq:fluct-diss}, we rewrite the heating rate as
\barr
\dot{\mathcal{Q}}_\chi &\approx& \frac{\rho_s \rho_\chi}{M^2 (T_b/m_s)}  \mathcal{B}(V_{\chi b}; T_b/m_s) \times  \left(T_b - \frac{M}{m_\chi}T_\chi \right) \nonumber\\
&+& \frac{\rho_s \rho_\chi}{M (T_b/m_s)}  \frac{T_\chi}{m_\chi} V_{\chi b}^2 \mathcal{A}(V_{\chi b}; T_b/m_s).\label{eq:heat-narrow}
\earr
These expressions hold provided $T_\chi/m_\chi \ll V_{\chi b}^2$. This limiting case can then be sub-divided into two sub-cases:\\

$\bullet$ Either $T_b/m_s \gtrsim V_{\chi b}^2 \gg T_\chi/m_\chi$, in which case the variance of the relative thermal motion is $v_{\rm th}^2 \approx T_b/m_s$. Hence, Eq.~\eqref{eq:momentum-narrow} becomes identical to Eq.~\eqref{eq:momentum-MB} obtained under the assumption of a MB distribution. Neglecting $m_s T_\chi/(m_\chi T_b)$ in the first term of Eq.~\eqref{eq:heat-narrow}, we see that this equation becomes identical to Eq.~\eqref{eq:heat-MB}.\\

$\bullet$ Or $T_b/m_s \ll V_{\chi b}^2$, in which case we also have $V_{\chi b} \gg v_{\rm th}$. Then, from Eqs.~\eqref{eq:A-asympt} and \eqref{eq:B-asympt}, $\mathcal{B}(V_{\chi b}) \approx  V_{\chi b}^2 \mathcal{A}(V_{\chi b})\approx V_{\chi b}^3 \overline{\sigma}(V_{\chi b})$, independent of temperature, and valid whether the coefficients are evaluated at $T_b/m_s$ or $v_{\rm th}$. In that case again, Eqs.~\eqref{eq:momentum-narrow} and \eqref{eq:momentum-MB} are identical, and so are Eqs.~\eqref{eq:heat-narrow} and \eqref{eq:heat-MB}, which both reduce to $\dot{\mathcal{Q}}_{\chi} \approx (\rho_s \rho_\chi m_s/M^2) V_{\chi b}^3 \overline{\sigma}(V_{\chi b}) $.

\subsubsection{Case where $T_\chi/m_\chi \ll T_b/m_s$ and $V_{\chi b}^2 \ll T_b/m_s$}

If the support of $f_\chi$ is well within the characteristic scale of the coefficients $\mathcal{A}$ and $\mathcal{B}$, i.e.~if both $V_{\chi b}^2$ and $T_\chi/m_\chi$ are much smaller than $T_b/m_s$, then, regardless of the relative ordering of $V_{\chi b}^2$ and $T_\chi/m_\chi$, we can again Taylor-expand $\mathcal{A}$ and $\mathcal{B}$ near $w = 0$. The momentum and heating rate are given by Eq.~\eqref{eq:momentum-narrow} and \eqref{eq:heat-narrow} evaluated at $w = 0$ and $T_b/m_s \approx v_{\rm th}^2$. Here again, we may neglect $m_s T_\chi/(m_\chi T_b)$ in the first term of Eq.~\eqref{eq:heat-narrow}, hence find that Eqs.~\eqref{eq:momentum-narrow} and \eqref{eq:heat-narrow} are identical to Eqs.~\eqref{eq:momentum-MB} and \eqref{eq:heat-MB}, respectively.

To conclude, we have identified two limiting cases where Eqs.~\eqref{eq:momentum-MB} and \eqref{eq:heat-MB} still hold, even if the underlying DM distribution is not MB:
\barr
T_\chi/m_\chi &\ll& V_{\chi b}^2 \nonumber \\
&\textrm{or}&  \label{eq:narrow-cond}\\
V_{\chi b}^2 \ll T_b/m_s \ \ \ &\textrm{and}& \ \ \ T_\chi/m_\chi \ll T_b/m_s . \nonumber 
\earr
In these limiting cases, the momentum and heat-exchange rate become closed equations for the DM mean velocity $\bs{V}_\chi$ and its \emph{effective} temperature $T_\chi$, regardless of the underlying velocity distribution, provided it is sufficiently narrow. 

\section{Fokker-Planck approximation to the collision operator} \label{sec:FP}

%
%

\subsection{Motivation}
If the DM distribution is not set to MB by self-interactions, we must explicitly solve for $f_\chi$ to get the momentum and heat exchange rates when $f_\chi$ is too broad. Specifically, the closed-form expressions derived in Section \ref{sec:closed-form} no longer apply when \eqref{eq:narrow-cond} is not satisfied, i.e.~if
\barr
 V_{\chi b}^2 & \lesssim& T_\chi/m_\chi \  \ \ \textrm{and} \ \ \  T_b/m_s \lesssim T_\chi/m_\chi. \label{eq:FP-validity}
\earr
Conveniently, this corresponds to the regime where the characteristic change in the DM velocity per scattering event is smaller than the variance of the DM distribution. Indeed, during an elastic scattering event $\chi + s \rightarrow \chi+ s$, the DM velocity changes by
\beq
\bs{v}_\chi' - \bs{v}_\chi = \frac{m_s}{M} v_{\chi s} (\hat{n}' - \hat{n}),  \label{eq:Delta-v}
\eeq
This implies
\barr
\frac{(\Delta \bs{v})^2}{T_\chi/m_\chi} \sim \left(\frac{m_s}{M}\right)^2 \left(1 + \frac{V_{\chi b}^2 + T_b/m_s}{T_\chi/m_\chi}\right) \lesssim 1. \label{eq:Dv-over-v}
\earr
This is precisely the regime where the integral collision operator \eqref{eq:C_pchi} can be approximated by a diffusion or Fokker-Planck (FP) operator. Such a collision operator is more amenable to efficient numerical evolution than the full integral operator \eqref{eq:C_pchi}. Thinking of this approximation in terms of a discretized distribution function, the exact integral collision operator corresponds to a multiplication by a full matrix, while the FP operator corresponds to a tri-diagonal matrix.

Strictly speaking, the diffusion approximation is expected to be most accurate when the velocity change per scattering is \emph{much} smaller than the width of the DM distribution, i.e., from Eq.~\eqref{eq:Dv-over-v}, when $m_s \ll m_\chi$. When this condition is not satisfied, there isn't a well-defined small parameter in which to expand the integral collision operator, see e.g.~\cite{Binder_16}.

To minimize errors even when scattering is not strictly diffusive, we will construct the FP collision operator with a ``top-down" approach. Rather than starting from the exact collision operator and Taylor-expanding in $\Delta v$ (as done, e.g.~in studies of kinetic decoupling of neutralino DM  \cite{Jungman_96} scattering with relativistic leptons \cite{Bertschinger_06, Bringmann_07, Bringmann_09}), we start from a general second-order diffusion operator, and uniquely determine all coefficients by demanding that this operator satisfies essential properties of the exact collision operator. Specifically, we enforce that the operator conserves DM number, satisfies detailed balance, and gives the correct momentum and heat exchange rates for a given distribution $f_\chi$. This ensures that we can use the FP operator through the entire evolution of the DM distribution function (with a subtle caveat, which we discuss later on). Indeed, even if the DM distribution is too narrow for the diffusion approximation to strictly apply, our FP operator is constructed in such a way to give the correct closed-form expressions for the momentum and heat-exchange rates that apply in this case. It therefore leads to the correct evolution of the bulk momentum and effective temperature, as well as the correct heat and momentum-exchange rates, even if the DM distribution is inaccurately computed. 

In summary, either the DM velocity distribution is broad, in which case the FP approximation is accurate, or it is narrow, in which case our top-down FP operator still recovers the correct heat and momentum-exchange rates. We defer to future work a quantitative study of the accuracy of this approximation in the intermediate regime.

\subsection{Top-down construction of the FP operator}

We seek to approximate the exact collision operator by a second-order differential operator of the form \cite{Thorne_Blandford}
\barr
C[f_\chi](\bs{v}) = \frac{\partial}{\partial v^i} \left\{ \frac12 \frac{\partial}{\partial v^j}\left[D^{ij}(\bs{v}) f_\chi(\bs{v})\right] + d^i(\bs{v})f_\chi (\bs{v}) \right\}.~~~~~ \label{eq:FP-general}
\earr
This operator explicitly conserves particle number, i.e. satisfies Eq.~\eqref{eq:number}. We now determine the vector $\bs{d}$ and symmetric tensor $D^{ij}$ so that this approximate collision operator gives the correct momentum and heat exchange rates, and satisfies detailed balance.

Multiplying Eq.~\eqref{eq:FP-general} by $ m_\chi \bs{v}$ and integrating over velocities, we find, after integrating by parts,
\beq
\dot{\bs{\mathcal{P}}}_{\chi} = - \rho_\chi \int d^3 v ~ \bs{d}(\bs{v}) f_\chi(\bs{v}). \label{eq:dVdt-FP}
\eeq 
For this expression to be identical to Eq.~\eqref{eq:momentum2a} \emph{for any distribution} $ f_\chi$, it must be that the vector $\bs{d}(\bs{v})$ is the opposite of the velocity drift vector:
\beq
\bs{d}(\bs{v}) \equiv  - \frac{d \langle \Delta \bs{v} \rangle}{dt}\big{|}_{\bs{v}} = \frac{\rho_s}{M} \mathcal{A}(w; T_b/m_s) \bs{w}.\label{eq:d}
\eeq
Similarly, multiplying Eq.~\eqref{eq:FP-general} by $\frac12 (\bs{v} - \bs{V}_\chi)^2$ and integrating over velocities, we find the heating rate
\barr
\dot{\mathcal{Q}}_{\chi} &=& \frac12 \rho_\chi \int d^3 v \left[ D^{ii} - 2 (\bs{v} - \bs{V}_\chi) \cdot \bs{d}(\bs{v})\right] f_\chi(\bs{v}). \label{eq:dTdt-FP}
\earr
For this expression to be identical to Eq.~\eqref{eq:heat2a} \emph{for any distribution} $ f_\chi$, it must be that the trace of $D^{ii}$ is the trace of the velocity diffusion tensor:
\barr
D \equiv D^{ii} \equiv \frac{d\langle (\Delta \bs{v})^2 \rangle}{dt}\big{|}_{\bs{v}} = 2 \frac{m_s \rho_s}{M^2}  \mathcal{B}(w; T_b/m_s). \label{eq:D}
\earr
To determine the anisotropic components of $D^{ij}$, we enforce detailed balance by imposing that the number flux -- the quantity in brackets in Eq.~\eqref{eq:FP-general} -- vanishes for a MB distribution at temperature $T_b$, and mean velocity $\bs{V}_b$ (with mass $m_\chi$). This condition is obtained by integrating Eq.~\eqref{eq:FP-general} over any finite velocity volume and using Stokes' theorem. This constrains the FP coefficients to satisfy
\beq
\partial_j D^{ij} - \frac{m_\chi}{T_b} (v - V_b)_j D^{ij}  + 2 d^i = 0. \label{eq:db}
\eeq
Isotropy in the rest-frame of the scatterers imply that the tensor $D^{ij}$ only has two independent components, longitudinal and transverse to $\bs{w} = \bs{v} - \bs{V}_b$, respectively, and whose magnitude only depend on $w$:
\barr
&&D^{ij}(\bs{v}) = D_{||}(w) \hat{w}^i \hat{w}^j + D_{\bot}(w) \left(\delta^{ij} - \hat{w}^i \hat{w}^j\right). \nonumber \\
&& =\frac{D_{||}(w)}{2} \left(3 \hat{w}^i \hat{w}^j - \delta^{ij} \right) + \frac{D(w)}2 \left(\delta^{ij} - \hat{w}^i \hat{w}^j\right). \label{eq:Dij-geometry}
\earr
Inserting this expression into Eq.~\eqref{eq:db}, we find that $D_{||}(w)$ satisfies the following first-order differential equation (see \cite{YAH_09} for equivalent equations in a very different context):
\barr
&& w \frac{d D_{||}}{d w} + 3 D_{||}(w) - \frac{m_\chi w^2}{T_b} D_{||} = D - 2 \bs{w} \cdot \bs{d}\nonumber\\
&&= 2 \frac{m_s}{M}\frac{\rho_s}{M} \left(\mathcal{B}(w; T_b/m_s) - \frac{M}{m_s} w^2 \mathcal{A}(w; T_b/m_s) \right). \label{eq:db-D||}
\earr
Using the fluctuation-dissipation relation \eqref{eq:fluct-diss}, we find
\beq
D_{||}(w) = 2 \frac{\rho_s T_b}{M^2} \mathcal{A}(w; T_b/m_s).
\eeq
We have therefore entirely determined the coefficients $D^{ij}$ and $\bs{d}$ and uniquely defined the FP operator.

To summarize, using Eq.~\eqref{eq:db}, the FP operator is
\barr
\boxed{C[f_\chi](\bs{v}) = \frac12 \frac{\partial}{\partial v^i} \left\{ D^{ij} \left(\frac{\partial f_\chi}{\partial v^j} + \frac{m_\chi}{T_b} (v - V_b) _j f_\chi \right) \right\}},~~~~~ \label{eq:FP}
\earr
where the tensor $D^{ij}$ is given by 
\begin{empheq}[box=\fbox]{align}
D^{ij}(\bs{v}) &= \frac{\rho_s m_s}{M^2} \Big{\{} \mathcal{B}(w; T_b/m_s)\left(\delta^{ij} - \hat{w}^i \hat{w}^j\right)   \nonumber\\
& ~~~~~~~ + \frac{T_b}{m_s} \mathcal{A}(w; T_b/m_s)  \left(3 \hat{w}^i \hat{w}^j - \delta^{ij} \right) \Big{\}}.
\end{empheq}
The FP operator \eqref{eq:FP} is an approximation to the full collision operator, expected to be most accurate when scattering events typically change the DM velocity by a small fractional amount. It conserves the DM number, and gives the correct rates of momentum and heat exchange, at a given distribution function. Finally, it satisfies detailed balance, i.e.~preserves the Maxwell-Boltzmann distribution at temperature $T_b$ and mean velocity $\bs{V}_b$. In this sense, it is the \emph{best possible diffusion approximation to the full collision operator}, even when velocity changes per scattering are not small. It most notably differs from previously derived expressions (e.g.~\cite{Binder_16} and references therein) by the velocity dependence and anisotropy of the diffusion tensor.

\section{Background evolution} \label{sec:background}

\subsection{Homogeneous FP equation}

We first study the evolution of the background homogeneous and isotropic distribution $\overline{f}_\chi$, in the absence of perturbations. Setting $\bs{V}_b = 0$, the FP operator \eqref{eq:FP} becomes
\barr
\overline{C}[\overline{f}_\chi](v) \equiv \frac{\rho_s T_b}{M^2 v^2} \frac{\partial}{\partial v}\left\{ v^2 \mathcal{A}(v) \left(\frac{\partial \overline{f}_\chi}{\partial v} + \frac{m_\chi}{T_b} v \overline{f}_\chi \right) \right\},~~~~ \label{eq:FP-bg}
\earr
where all densities and temperatures are background (homogeneous) quantities, and the coefficient $\mathcal{A}$ is to be evaluated at $T_b/m_s$. The Boltzmann equation in the expanding background is 
\barr
\frac{d }{dt}(a^{-3}\overline{f}_\chi)  \equiv \left(\partial_t - H v \partial_v \right)(a^{-3} \overline{f}_\chi) =  a^{-3} \overline{C}[\overline{f}_\chi](v), ~~\label{eq:Boltzmann-bg}
\earr
where $H$ is the Hubble rate and we used $\overline{n}_\chi \propto a^{-3}$. The evolution of the effective temperature is then
\barr
&&\frac1{a^2} \frac{d}{dt}(a^2 T_\chi) = \frac{m_\chi}{3} \int d^3 v ~ v^2~ \overline{C}[\overline{f}_\chi](v) \nonumber\\
&&=  - \frac{2 \overline{\rho}_s T_b m_\chi}{3M^2 } \int d^3 v ~  \mathcal{A} (v) \left(v \partial_v \overline{f}_\chi + \frac{m_\chi v^2}{T_b} \overline{f}_\chi \right), \label{eq:dTdt-FP-bg}
\earr
where we have integrated by parts to get the second expression.

\subsection{Condition of validity of the MB distribution}

We now show that the MB distribution with temperature $T_\chi$ is the solution of the \emph{background} FP equation \emph{if and only if} the diffusion coefficient $\mathcal{A}$ is independent of velocity.

We start by defining a MB temperature $T_\chi^{\rm MB}$, whose evolution is obtained by substituting a MB distribution $f_\chi^{\rm MB}$ with temperature $T_\chi^{\rm MB}$ in Eq.~\eqref{eq:dTdt-FP-bg}. This is what is usually used as the DM temperature in the existing literature, and in general is not equal to the correct effective temperature defined in Eq.~\eqref{eq:Tchi}. We find
\beq
\frac1{a^2} \frac{d}{dt}\left(a^2 T_\chi^{\rm MB}\right) =  \frac{2 \overline{\rho}_s m_\chi}{M^2} \mathcal{A}^{\rm MB} \left(T_b - T_\chi^{\rm MB}\right), \label{eq:dTdt-MB-bg}
\eeq
where the velocity-independent coefficient $\mathcal{A}^{\rm MB}$ is the following weighted average of $\mathcal{A}(v; T_b/m_s)$:
\beq
\mathcal{A}^{\rm MB} \equiv  \int d^3 v  \frac{m_\chi v^2}{3 T_\chi^{\rm MB}}  ~f_\chi^{\rm MB}(v; T_\chi^{\rm MB}) \mathcal{A}(v; T_b/m_s).~~~  \label{eq:AMB}
\eeq
This expression can be independently derived from Eq.~\eqref{eq:heat-MB}, setting $V_{\chi b} = 0$:
\barr
\frac1{a^2} \frac{d}{dt}\left(a^2 T_\chi^{\rm MB}\right) &=& \frac{2 \overline{\rho}_s m_\chi}{3 M^2 v_{\rm th}^2} \mathcal{B}(0; v_{\rm th}^2) \left(T_b - T_\chi^{\rm MB}\right)\nonumber\\
&=&\frac{2 \overline{\rho}_s m_\chi}{M^2} \mathcal{A}(0; v_{\rm th}^2) \left(T_b - T_\chi^{\rm MB}\right)
\earr
where $v_{\rm th}^2$ is given by Eq.~\eqref{eq:vth}, with $T_\chi \rightarrow T_\chi^{\rm MB}$, and we used the fluctuation-dissipation relation \eqref{eq:fluct-diss-0}. We conclude that 
\beq
\mathcal{A}^{\rm MB} = \mathcal{A}(0; T_b/m_s + T_\chi^{\rm MB}/m_\chi).
\eeq 
We now define $C^{\rm MB}$ to be the FP operator obtained by replacing $\mathcal{A}(v)$ by $\mathcal{A}^{\rm MB}$ in Eq.~\eqref{eq:FP-bg}. Let us show that the MB distribution at temperature $T_\chi^{\rm MB}$ is a solution of 
\beq
a^3 \frac{d (a^{-3} f_\chi^{\rm MB})}{dt} = C^{\rm MB}[f_\chi^{\rm MB}]. \label{eq:Boltzmann-MB}
\eeq
The left-hand-side is 
\beq
a^3 \frac{d (a^{-3} f_\chi^{\rm MB})}{dt} = \frac32 \frac{d}{dt} \ln(a^2 T_\chi^{\rm MB}) \left( \frac{m_\chi v^2}{3 T_\chi^{\rm MB}} - 1 \right)  f_\chi^{\rm MB}(v). \label{eq:FP-MB}
\eeq
On the other hand, since $\mathcal{A}^{\rm MB}$ is independent of $v$, we may rewrite the right-hand-side as
\barr
C^{\rm MB}[f_\chi^{\rm MB}](v) &=& \frac{3 m_\chi \overline{\rho}_s}{M^2 T_\chi^{\rm MB}} \mathcal{A}^{\rm MB} (T_b - T_\chi^{\rm MB})  \nonumber\\
&& \times \left( \frac{m_\chi v^2}{3 T_\chi^{\rm MB}} - 1 \right) f_\chi^{\rm MB}(v).
\earr
We see that Eq.~\eqref{eq:Boltzmann-MB} is indeed satisfied for all $v$ provided $T_\chi^{\rm MB}$ is a solution of Eq.~\eqref{eq:dTdt-MB-bg}.

Let us now define $\Delta \mathcal{A}(v) \equiv \mathcal{A}(v) - \mathcal{A}^{\rm MB}$, $\Delta C[\overline{f}_\chi] \equiv \overline{C}[\overline{f}_\chi] - C^{\rm MB}[\overline{f}_\chi]$ and $\Delta \overline{f}_\chi = \overline{f}_\chi - f_\chi^{\rm MB}$. Using Eq.~\eqref{eq:Boltzmann-MB}, the Boltzmann equation can be rewritten as the following inhomogeneous partial differential equation for $\Delta \overline{f}_\chi$:
\barr
&&a^3 \frac{d  (a^{-3} \Delta \overline{f}_\chi)}{dt} - \overline{C}[\Delta \overline{f}_\chi] = \Delta C[f_\chi^{\rm MB}]\nonumber\\
&&~~~~~ = \frac{m_\chi \overline{\rho}_s (T_\chi^{\rm MB} - T_b)}{M^2 T_\chi^{\rm MB} v^2} \frac{\partial}{\partial v} \left\{ v^3 \Delta \mathcal{A}(v) f_\chi^{\rm MB}(v) \right\}. \label{eq:DFP}
\earr
This equation does not admit $\Delta \overline{f}_\chi = 0$ as a solution, unless the right-hand-side vanishes for all $v$, which is the case if and only if $\mathcal{A} = \mathcal{A}^{\rm MB}$ is independent of velocity. We therefore conclude that the MB distribution is a solution of the homogeneous Boltzmann-FP equation if and only if $\mathcal{A}$ is independent of velocity. As a consequence, the temperature $T_\chi^{\rm MB}$ is in general not equal to the effective temperature $T_\chi$ defined in Eq.~\eqref{eq:Tchi}. More importantly, the background heat exchange rate is not correctly captured by the MB approximation if $\mathcal{A}$ is velocity dependent.

\subsection{Numerical solution during radiation domination}

For most cosmological probes of interest, thermal and kinematic decoupling of the DM occurs deep in the radiation era given current upper bounds, and we focus on this epoch here to simplify the problem and gain intuition. We therefore assume that $H(a) = H_0 \Omega_r^{1/2} a^{-2}$. We also assume that DM scatters off a single species $s$, with background density $n_s = n_{s0} a^{-3}$. The generalization to an arbitrary Hubble expansion rate and to multiple scatterers is straightforward, but does not allow to write equations in the simple dimensionless form that we now derive.

\subsubsection{Thermal approximation}

We assume that, up to very small perturbations, heat exchange with the DM does not noticeably affect the standard temperature evolution of the photon-baryon plasma, which is in thermal equilibrium at these redshifts, with a temperature $T_b = T_0 /a$. We specialize to power-law cross sections of the form \eqref{eq:sigma_n}, with $n > -3$, so that the DM is thermally coupled to the baryons at early times, and eventually decouples. We denote by $a_n$ the characteristic scale factor at thermal decoupling \cite{ACK15}:
\beq
a_n^{\frac{n+3}{2}} \equiv 2 c_n \sigma_n \frac{n_{s0}}{H_0 \Omega_r^{1/2}} \left(\frac{M^2}{m_s m_\chi}\right)^{\frac{n-1}{2}} \left(\frac{T_0}{M}\right)^{\frac{n+1}{2}}, \label{eq:a_n}
\eeq
where $c_n$ was defined in Eq.~\eqref{eq:cn}.
We define the rescaled scale factor $y$ and dimensionless effective temperature $X$:
\beq
y \equiv a/a_n, \ \ \ \ \ X \equiv T_\chi/T_b.
\eeq
We also define the dimensionless heat exchange rate
\beq
Q \equiv \frac{\dot{\mathcal{Q}}_\chi}{\frac32 \overline{n}_\chi H T_b} = \frac{d}{dy}(y X).\label{eq:Q-dimless}
\eeq
In terms of these variables, Eq.~\eqref{eq:dTdt-MB-bg} may be rewritten as 
\barr
\frac{d (y X^{\rm MB})}{d y} = \left(\frac{1 + (m_s/m_\chi) X^{\rm MB}}{1 + m_s/m_\chi} \right)^{\frac{n+1}2}\frac{1 - X^{\rm MB}}{y^{\frac{n+3}2}}.~~ \label{eq:temp-num}
\earr
This makes it clear that, for a given $n$, the evolution of $T_\chi^{\rm MB}/T_b$ is a function of $y$ and $m_s/m_\chi$ only. 

\begin{figure}
\includegraphics[width = \columnwidth]{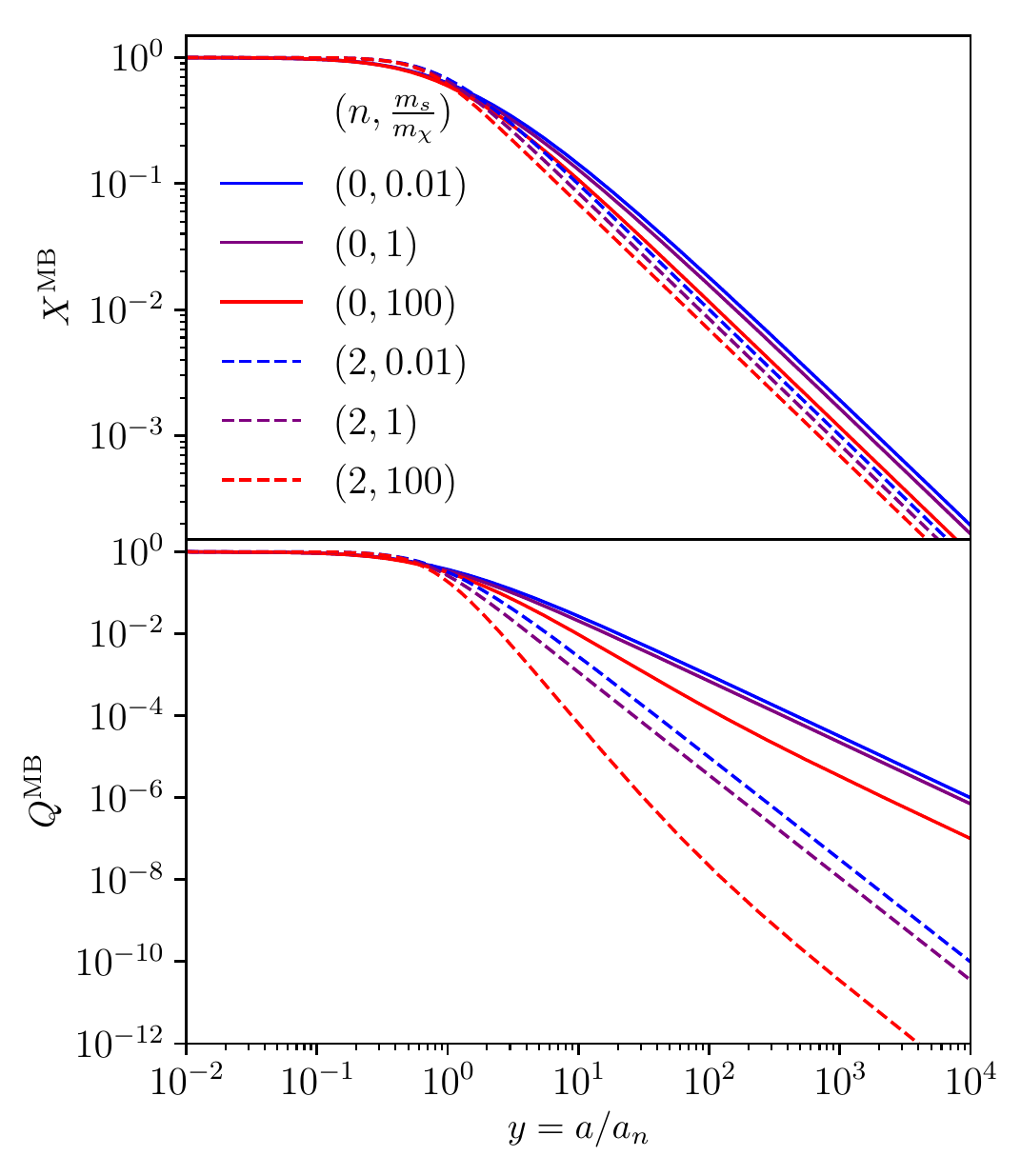}
\caption{Evolution of the rescaled DM temperature $X \equiv T_\chi/T_b$ and dimensionless heat-exchange rate $Q \equiv \dot{\mathcal{Q}}_\chi/\frac32 n_\chi H T_b$ as a function of the rescaled scale factor $y = a/a_n$, in the Maxwell-Boltzmann approximation. }
\label{fig:temperature}
\end{figure}

We show the numerical solution of Eq.~\eqref{eq:temp-num} and resulting dimensionless heat-exchange rate in Fig.~\ref{fig:temperature}, for $n = 0$ and 2, and for $m_s/m_\chi = 0.01, 1, 100$. We can understand its asymptotic features as follows.

First, for $y \ll 1$, we have $X^{\rm MB} \rightarrow 1$ and $Q^{\rm MB} \rightarrow 1$. More precisely, we find the following asymptotic expansion:
\beq
X^{\rm MB} \approx 1 - y^{\frac{n+3}{2}} + \frac{n + 5 + 4 m_s/m_\chi}{2(1 + m_s/m_\chi)} y^{n+3},
\eeq
valid up to corrections of order $\mathcal{O}\left(y^{\frac32 (n+3)}\right)$. We use this expression to initialize our numerical integration of Eq.~\eqref{eq:temp-num}.

Secondly, for $y \gtrsim 1$, we have $X^{\rm MB} \sim 1/y$. Once $(m_s/m_\chi) X^{\rm MB} \ll 1$, we find
\beq
Q^{\rm MB} \approx \frac{y^{-\frac{n+3}{2}}}{(1 + m_s/m_\chi)^{\frac{n+1}2}}, \ \ \ \ y \gg \max(1, m_s/m_\chi).
\eeq
If $m_s/m_\chi \gg 1$, there is moreover an intermediate regime,
\beq
Q^{\rm MB} \sim \frac1{y^{n+2}},  \ \ \  1 \ll y \ll m_s/m_\chi.
\eeq

\subsubsection{Fokker-Planck solution}

We rewrite the Boltzmann equation in terms of $y = a/a_n$ and dimensionless variables 
\beq
x \equiv \sqrt{\frac{m_\chi}{T_b}} ~v, \ \ \ \ \ \ \ \mathcal{N}(y,x) \equiv 4 \pi v^3 \overline{f}_\chi(a, v).
\eeq
The function $\mathcal{N}$ is the distribution of velocities per logarithmic interval, such that $\int d\ln x~ \mathcal{N} = 1$. The Boltzmann equation becomes
\barr
&& y \partial_y \mathcal{N} - \frac12 x \partial_x \mathcal{N} = \frac12 \left(\frac{m_\chi}{M}\right)^{\frac{n+1}{2}} y^{-\frac{n+3}{2}} \times \nonumber\\
&&   x\partial_x \left\{ x^2 \alpha_n\left(x\sqrt{\frac{m_s}{m_\chi}}\right)\left[ \partial_x (\mathcal{N}/x^3) + x (\mathcal{N}/x^3)\right] \right\}.~~~~~\label{eq:N-num}
\earr
Again, the evolution of $\mathcal{N}(y, x)$ only depends on the index $n$ and on the mass ratio $m_s/m_\chi$. The rescaled effective temperature $X = T_\chi/T_b$ is then obtained from
\beq
X = \frac13 \int d\ln x ~ x^2 \mathcal{N},\label{eq:X}
\eeq
and the dimensionless heat-exchange rate $Q$ is again obtained from $Q = d(yX)/dy$. 

We evolve Eq.~\eqref{eq:N-num} numerically using a method similar in spirit to that of Refs.~\cite{Hirata_09, YAH_11}: we discretize the collision operator with a tridiagonal matrix, which exactly satisfies detailed balance and conserves DM number; we use a fixed logarithmic timestep $\Delta \ln y$ and set $\Delta \ln x = \Delta \ln y/2$ so that Hubble redshifting can be exactly accounted for by simply moving down the distribution $\mathcal{N}$ by one bin at each timestep. As a sanity check, we verified that our code does recover a MB distribution with temperature $T_\chi^{\rm MB}$ when solving Eq.~\eqref{eq:N-num} with the velocity-averaged diffusion coefficient $\alpha_n^{\rm MB} \equiv (1 + (m_s/m_\chi) X)^{(n+1)/2}$. For increased accuracy, we solve for $\Delta \mathcal{N} \equiv \mathcal{N} - \mathcal{N}^{\rm MB}$, as this vanishes initially, and is less prone to numerical errors. 

We find that the amplitude of the distortion from MB and the fractional change in heating rate $|Q - Q^{\rm MB}|/Q$ both increase with $m_s/m_\chi$, as well as with $|n+1|$. For $m_s /m_\chi \leq 1$, we find that the MB approximation reproduces the heat-exchange rate to better than 15\% accuracy for $-2 \leq n \leq 4$. We emphasize that, by no means does this imply that perturbed quantities are recovered this accurately by the MB approximation. 

\begin{figure*}
\includegraphics[width = 2\columnwidth]{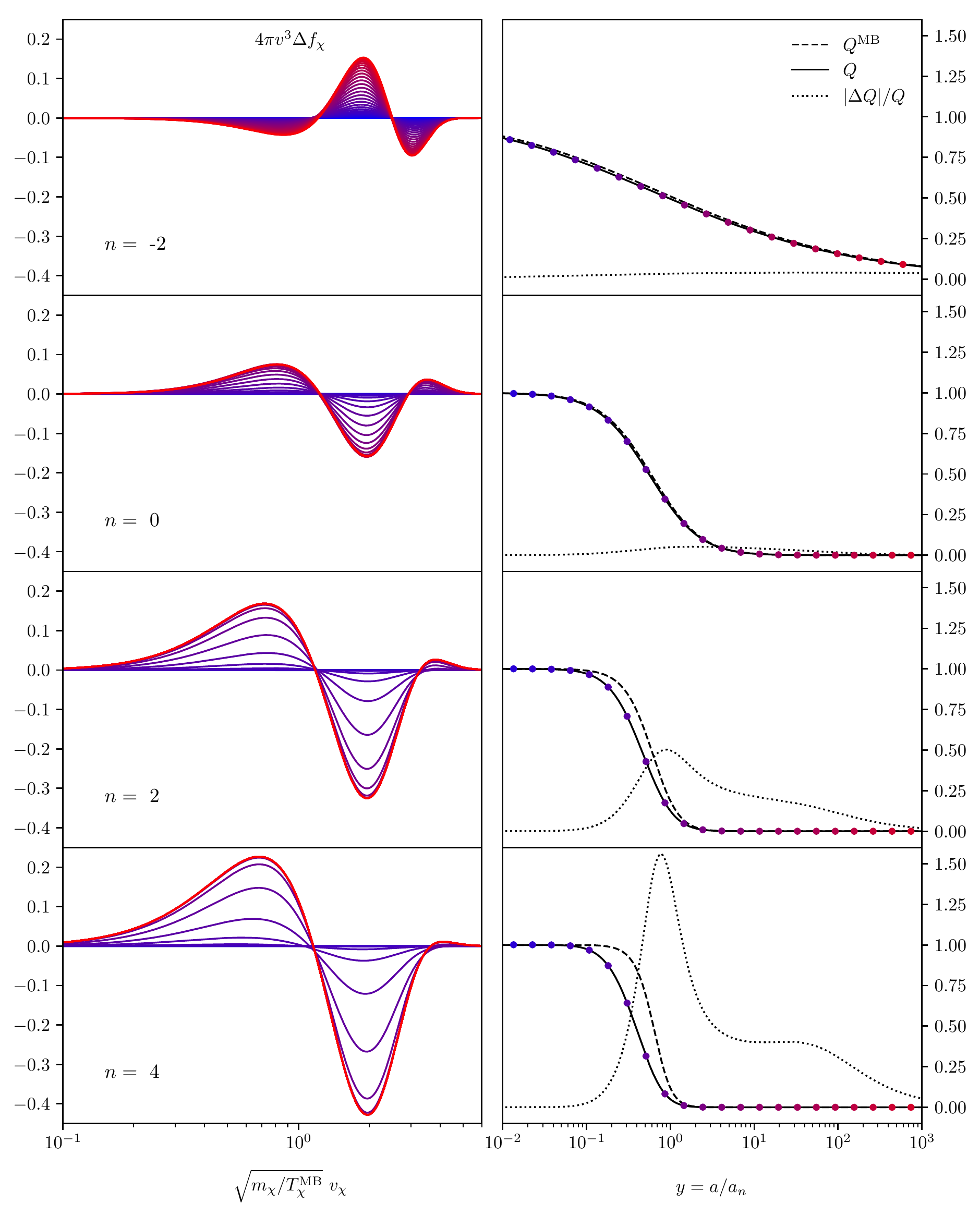}
\caption{For a baryon-to-DM mass ratio $m_s/m_\chi = 100$, the left column shows the departure of the DM velocity distribution from MB, for several values of the index $n$ such that $\sigma_{\chi b} \propto v_{\chi b}^n$. The right column shows the corresponding dimensionless heat-exchange rates and their fractional difference, as a function of $a/a_n$, where $a_n$ is the characteristic scale factor of thermal decoupling. The colored points on the right column correspond to the times at which the distortion is plotted on the left. We see that the characteristic distortion amplitude and $|\Delta Q|/Q$ increase with $|n+1|$. For $n = 2$ and, especially, $ n = 4$, the heat exchange rates computed in the MB approximation and with the Boltzmann-Fokker-Planck equation differ by order unity.}
\label{fig:distortion_100}
\end{figure*}

\begin{figure}
\includegraphics[width = \columnwidth]{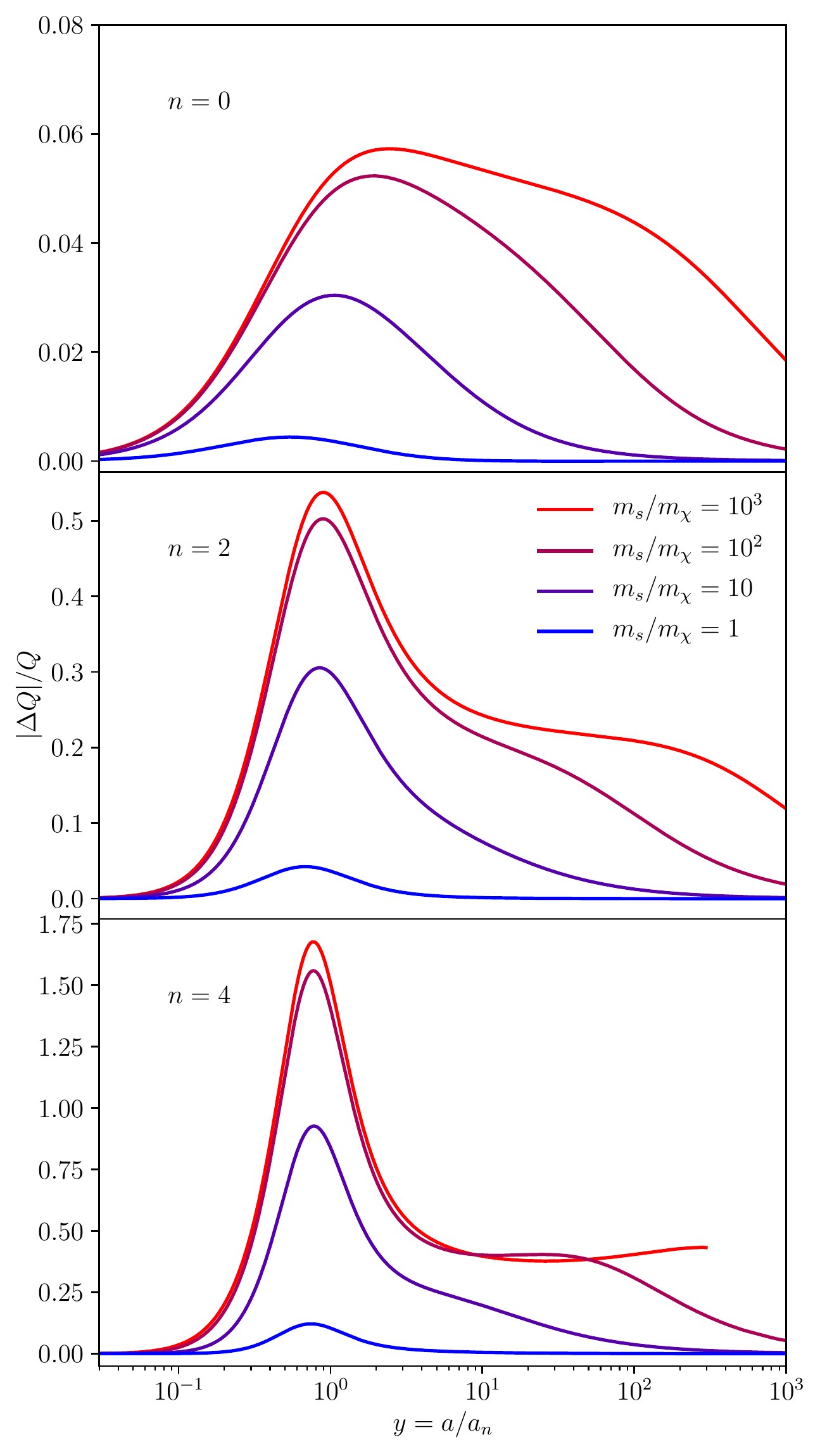}
\caption{Fractional difference in the homogeneous heat-exchange rate computed in the MB approximation and from the FP equation, as a function of rescaled scale factor, for $n = 0, 2, 4$, and for baryon-to-DM mass ratios $m_s/m_\chi = 1, 10, 10^2, 10^3$, from bottom to top on each panel. For $n = 0$, the homogeneous exchange rate does not differ by more than a few percent, while for $n = 4$, there are order-unity differences for $m_s/m_\chi \gtrsim 10$. For $n = 4$, $m_s/m_\chi = 10^3$, the curve is truncated at $y =300$, beyond which $Q \lesssim 10^{-15}$.}\label{fig:DQ}
\end{figure}

For $m_s/m_\chi \gg 1$, however, the MB approximation and FP solution have order-unity differences, in particular for $n \geq 2$. This is illustrated in Fig.~\ref{fig:distortion_100}, where we show, for $m_s/m_\chi = 100$, the departure of the DM distribution from MB, and the resulting heat-exchange rates. Fig.~\ref{fig:DQ} shows the fractional difference in heat-exchange rate between the MB approximation and the FP solution, for several mass ratios and indices $n$. \newversion{We see that this difference is largest around the time of decoupling, but remains significant all the way to $y \sim m_s/m_\chi$ (corresponding to $T_\chi/m_\chi \sim T_b/m_s$), i.e.~until long after thermal decoupling if $m_s \gg m_\chi$. At later times, the heating rates eventually converge, despite the fact that non-thermal distortions to the DM distribution are largest (see left panel of Fig.~\ref{fig:distortion_100}). This is to be expected from our discussion in Section \ref{sec:closed-form}: once $T_\chi/m_\chi \ll T_b/m_s$, the heating rate takes on a closed form, independent of the detailed shape of $f_\chi$.}

It is noteworthy that, somewhat counterintuitively, the MB approximation happens to be most accurate for cases where decoupling takes the longest. The reason is that the swiftness of decoupling plays no role in the accuracy of the MB approximation: it is only dependent on the steepness of the diffusion coefficient across the width of the DM distribution. It is also worthwhile noticing that the MB approximation systematically overestimates the efficiency of heat exchange.


\section{Perturbations} \label{sec:pert}

\subsection{Invalidity of the MB approximation for perturbations}

In order to study the impact of DM-baryon scattering on the evolution of perturbation, hence CMB anisotropies and large-scale-structure observables, we must solve for the perturbed Boltzmann-Fokker-Planck equation. 

Let us first note that, even if the diffusion coefficient is isotropic and independent of velocity, the MB distribution is \emph{not} a solution of the Boltzmann equation as soon as there are spatial perturbations. To see this, consider the case where $D^{ij} = D_{||} \delta^{ij}$, where $D_{||}$ is independent of velocity. Insert the MB distribution into both sides of the perturbed Boltzmann equation. Equate the coefficients of $ 1- m_\chi (\bs{v} - \bs{V}_\chi)^2/3T_\chi^{\rm MB}$, and find the following equation for the temperature:
\beq
\frac1{a^2}\left(\partial_t + \frac1{a} \bs{v} \cdot \nabla_{\bs{x}} \right) (a^2 T_\chi^{\rm MB}) = \frac{m_\chi}{T_b} D_{||} (T_b - T_\chi^{\rm MB}).
\eeq
The advection term $\bs{v} \cdot \nabla_{\bs{x}}$ implies that the temperature must depend on the velocity of the non-relativistic DM, hence that a MB distribution is not permitted. This would not be the case had the DM been relativistic: in this case, an anisotropic but still thermal momentum distribution would be permitted -- it is the case, for instance, of CMB photons, whose spectrum is a blackbody with a direction-dependent temperature. 

We therefore expect factor-of-a-few differences between the thermal approximation and the FP equation when computing the momentum-exchange rate, even for a nearly velocity-independent diffusion rate. \newversion{From our study of the background heat-exchange rates in Section \ref{sec:background}, we expect that the fractional difference will be the largest around thermal decoupling at redshift $z_n = a_n^{-1}$, and remain significant until $z \sim (m_\chi/m_s) z_n$ for $m_\chi \ll m_s$. For $m_\chi = 1$ MeV, the 95\%-confidence upper limits to the DM-baryon cross section of Ref.~\cite{Boddy_18} translate to $z_n \approx (16,  8) \times 10^4$ for the operators $\mathcal{O}_7 (n = 2)$ and $\mathcal{O}_3 (n = 4)$, respectively. This implies that fractional differences in heat and momentum-exchange rates ought to be significant from $z\sim 10^5$ all the way down to $z \sim 10^2$, i.e.~throughout the epochs relevant to CMB anisotropies, from which these limits are derived. We therefore expect that the differences between the thermal approximation and the FP treatment} ought to affect upper bounds and forecasts at the factor-of-a-few level.

\subsection{Boltzmann-Fokker-Planck hierarchy}

The most sensible method to numerically solve for the perturbed DM distribution is to first expand it (in Fourier space) on the basis of Legendre polynomials, as is usually done in the study of photons or neutrinos \cite{Ma_95}. We define the perturbation $\delta f_\chi$ as
\beq
\delta f_\chi \equiv (1 + \delta_\chi) f_\chi - \overline{f}_\chi,
\eeq
where $\delta_\chi \equiv \delta n_\chi/\overline{n}_\chi$. For a given Fourier wavenumber $\bs{k}$, we then expand $\delta f_\chi$ as
\barr
\delta f_\chi(t, \bs{k}, \bs{v}) = \sum_{\ell} i^{\ell} (2 \ell +1) f_{\ell}(t, \bs{k}, v) P_{\ell}(\hat{k} \cdot \hat{v}).
\earr
To write the perturbed Boltzmann equation, we must first specify a gauge. We adopt the conformal Newtonian gauge \cite{Ma_95}, with metric 
\beq
ds^2 = a^2 [-( 1+ 2 \psi) d\tau^2 +  (1 - 2 \phi) d\bs{x}^2].
\eeq 
The time appearing in the Boltzmann equation is the proper time of comoving observers\footnote{This $t$ is  therefore not the usual FLRW time coordinate.}, $dt \equiv a(1 + \psi) d \tau$. The Boltzmann equation can then be rewritten in terms of conformal time $\tau$ as
\beq
a^3\frac{d}{d\tau}\left[a^{-3} (1+\delta_\chi) f_\chi \right] = a ( 1+ \psi) C[(1 + \delta_\chi) f_\chi].
\eeq
Let us first compute the right-hand-side to linear order in perturbations:
\barr
&& C[(1 + \delta_\chi) f_\chi] = C[\overline{f}_\chi] \nonumber\\
 &&+ \sum_{\ell}i^\ell (2 \ell+1) \overline{C}[f_{\ell}(t, \bs{k}, v) P_{\ell}(\hat{k} \cdot \hat{v})],~~
\earr
where $\overline{C}$ is the background collision operator, obtained for $\bs{V}_b = 0$ and homogeneous and isothermal scatterers. It is relatively straightforward to show that
\beq
\overline{C}[f_{\ell}(t, \bs{k}, v) P_{\ell}(\hat{k} \cdot \hat{v})] = \left(\overline{C}[f_\ell] - \frac{\ell(\ell +1)}{2} \frac{D_{\bot}}{v^2} f_\ell \right) P_{\ell}(\hat{k} \cdot \hat{v}).
\eeq
The term $C[\overline{f}_\chi]$ has perturbations due to inhomogeneities in the scatterers's density and temperature, as well as their peculiar velocity. The latter is the most delicate to compute, due to the dependence of the diffusion tensor on $\bs{V}_b$. It is best to first rewrite
\beq
\overline{f}_\chi(\bs{v}) = \overline{f}_\chi(\bs{v} - \bs{V}_b) + (\bs{V}_b \cdot \hat{v}) \partial_v \overline{f}_\chi, 
\eeq
to linear order in perturbations. We then get
\barr
&&C[\overline{f}_\chi](\bs{v}) = \frac1{2 w^2} \frac{\partial}{\partial w} \left[ w^2 D_{||} \left(\partial_w \overline{f}_\chi + \frac{m_\chi}{T_s} w \overline{f}_\chi \right)\right]_{\bs{w} = \bs{v} - \bs{V}_b} \nonumber\\
&&~~~~~~~~~~~~ + \overline{C}[(\bs{V}_b \cdot \hat{v}) \partial_v \overline{f}_\chi]\nonumber\\
&&~~~~~~~~~~~~ = (1 + \delta_s)\overline{C}[\overline{f}_\chi](v)+  \delta T_b \frac{\delta \overline{C}}{\delta T_b}[\overline{f}_\chi] \nonumber\\
&&~~~~~ + (\bs{V}_b\cdot \hat{v})\left[ \overline{C}[\partial_v \overline{f}_\chi] - \frac{D_{\bot}}{v^2} \partial_v \overline{f}_\chi -\partial_v \left(\overline{C}[\overline{f}_\chi]\right) \right], \label{eq:C-pert}
\earr
where $\delta_s \equiv \delta \rho_s/\overline{\rho}_s$. The last line of Eq.~\eqref{eq:C-pert} reduces to $\frac12 D_{||} \frac{m_\chi}{T_b} \bs{V}_b \cdot \partial_{\bs{v}} \overline{f}_\chi$ if the diffusion tensor is isotropic and independent of velocity, as it should. 

Let us now consider the left-hand-side of the Boltzmann equation. To evaluate the total derivative operator $d/d\tau$, we must first clarify the meaning of $\bs{v}$: it is the proper velocity measured by a comoving observer, i.e.~,up to corrections of order $v^2$, and to linear order in metric perturbations, 
\beq
v^i = v_i = (1 - \phi - \psi) \frac{d x^i}{d \tau}. 
\eeq
The geodesic equation then translates to 
\beq
\frac{d \bs{v}}{d \tau} = - a H \bs{v} +  \dot{\phi} ~\bs{v} - \partial_{\bs{x}} \psi.
\eeq
Hence, to linear order, the total derivative operator is 
\beq
d/d \tau = \partial_{\tau} + \bs{v} \cdot \partial_{\bs{x}} + (- a H \bs{v} + \dot{\phi} \bs{v} - \partial_{\bs{x}} \psi) \cdot \partial_{\bs{v}}, 
\eeq
where we neglected perturbed quantities multiplying $\partial_{\bs{x}}$, which only acts on perturbations.
%
Putting everything together, and moreover assuming $\bs{V}_b = i \hat{k} V_b$, as is the case for scalar modes,
we arrive at the following hierarchy of equations for the perturbations:
\barr
&& a^3 \partial_\tau(a^{-3} f_\ell) - a H v \partial_v f_\ell + \frac{k v}{2 \ell +1} \left[\ell f_{\ell -1} - (\ell +1) f_{\ell +1} \right] \nonumber\\
&& = a\left(\overline{C} - \frac{\ell(\ell + 1)}{2} \frac{D_{\bot}}{v^2}\right)\left[f_\ell + \frac13 V_b \partial_v \overline{f}_\chi \delta_{\ell 1} \right] \nonumber\\
&& + \left( a( \psi + \delta_s) \overline{C}[\overline{f}_\chi] +  a\delta T_b \frac{\delta \overline{C}}{\delta T_b}[\overline{f}_\chi]- \dot{\phi} v \partial_v \overline{f}_\chi\right)\delta_{\ell 0} \nonumber\\
&& +\frac13  \left( k \psi \partial_v \overline{f}_\chi - a V_b \partial_v \left(\overline{C}[\overline{f}_\chi]\right)\right)\delta_{\ell 1}. 
\earr
This hierarchy is similar to that satisfied by massive neutrinos \cite{Ma_95}, with an additional collision term. 

The DM hierarchy must be solved simultaneously with the evolution of baryon perturbations. Baryons are themselves always an ideal thermal fluid, described by their density, bulk velocity and temperature. The continuity equation is unchanged, but the momentum equation gets an additional contribution due to scattering with DM, given by Eq.~\eqref{eq:momentum2}. This can be rewritten in terms of the dipole part of the DM perturbation, $f_1$, as follows:
\barr
\overline{\rho}_b \dot{V}_b|_{\chi b} = \frac{\overline{\rho}_s \overline{\rho}_\chi}{M} \int d^3 v ~ v\mathcal{A}(v) \left( f_1(v) + \frac13 V_b \partial_v \overline{f}_\chi \right).~~~~~ 
\earr
We see that the full shapes of $f_1(v)$ and $\overline{f}_\chi$ are relevant to the momentum-exchange rate. Notice that it is $\overline{\rho}_b$ that appears in the left-hand-side, as all baryons rapidly share momentum, but only the mass density of scatterers, $\overline{\rho}_s$, that appears on the right-hand-side.

The baryon temperature $T_b = \overline{T}_b + \delta T_b$ must also be solved for simultaneously. In addition to adiabatic cooling and Compton heating, DM-baryon interactions lead to the following cooling rate, to linear order in perturbations:
\barr
\frac32 \overline{n}_b \dot{T}_b |_{\chi b} &=& \frac{\overline{\rho}_s \overline{\rho}_\chi}{M} \int d^3 v \left( \overline{f}_\chi(v) + f_0(v) \right) \nonumber\\
&&\times \left(\frac{m_s}{M} \mathcal{B}(v; T_b/m_s) - v^2 \mathcal{A}(v; T_b/m_s) \right).~~~~~
\earr
Perturbations to this equation\footnote{It is worthwhile pointing out that none of the linear-cosmology studies of DM-baryon interactions seem to have considered perturbations to the baryon and DM temperatures thus far.} arise from the monopole of the perturbed DM distribution $f_0$, as well as perturbations to $T_b$ inside $\mathcal{A}$ and $\mathcal{B}$. 

Finally, the system is closed with the Einstein field equations, sourcing the potentials $\phi$ and $\psi$, which also depend on integrals of the moments $f_{\ell \leq 2}$.

A particular case of the perturbed Boltzmann-FP equation for neutralino-DM was first studied in Ref.~\cite{Bertschinger_06}. In that case, the DM scatters off relativistic leptons, and the diffusion tensor is isotropic and independent of velocity. This allowed the author of \cite{Bertschinger_06} to further expand the DM distribution function on the basis of eigenfunctions of the FP operator (see also Ref.~\cite{Binder_16}, and application in Ref.~\cite{Kamada_17} with a truncated hierarchy that reduces to ideal thermal fluid equations). In contrast, no analytic eigenfunctions are available for the velocity-dependent diffusion tensor that we first study here. One must therefore solve a system of partial differential equations for the $f_\ell(\tau, \bs{k}, v)$ rather than ordinary differential equations for the coefficients of the eigenfunctions. Still, the numerical burden should be manageable, and not much larger than what is required to solve the Boltzmann hierarchy for massive neutrinos. We defer to future work the implementation of the above Boltzmann-FP hierarchy into a CMB Boltzmann code, as well as the subsequent extraction of limits on DM-baryon cross sections from CMB anisotropy data.

\section{Limitations and extensions} \label{sec:late-coupling}

\subsection{Late-coupling scenarios}

So far we have only considered velocity dependences such that the DM starts thermally and kinematically coupled to baryons, and eventually decouples. For steep enough negative power laws of velocity, the reverse happens: the DM starts thermally and kinematically decoupled, and couples to baryons at late time. 

A concrete example is a Coulomb-like DM-baryon interaction, with cross section $\sigma \propto 1/v^4$, which could arise for instance for a milli-charged DM particle. Because of the late-time coupling in these scenarios, this is potentially detectable in 21-cm tomography \cite{Tashiro_14, Munoz_15}. It has recently been invoked \cite{Barkana_18, Fialkov_18} as a possible explanation of the EDGES 21-cm measurement \cite{Bowman_18}, though this would require the milli-charged particle to only make a small fraction of the total DM \cite{Munoz_18, Barkana_18b, Kovetz_18, Munoz_18b}. 

The late thermal decoupling in these scenarios leads to some modeling challenges. First of all, the bulk relative velocity between DM and baryons becomes larger than the baryon thermal motion for $z \lesssim 10^4$ (see Fig.~1 of Ref.~\cite{Dvorkin_14}), at least if the former is computed while neglecting the effect of interactions \cite{Boddy_18b}. As a consequence, relative motions enter the collision term \emph{non-perturbatively}, and one cannot linearize the momentum equation. An approximate treatment of this fundamentally non-linear effect was proposed in \cite{Dvorkin_14} and improved upon in \cite{Boddy_18b}, in both cases within the ideal thermal fluid approximation. It is not clear how accurate those approximations are, as no reference exact numerical solution exists to date. It is also not obvious how they ought to be implemented within the Boltzmann-FP formalism developed in this work.

In addition, if the DM does not strongly self-interact, the FP approximation (and, a fortiori, the thermal approximation) need not be accurate for such late-time coupling scenarios. To understand this, let us first recall the reason why the FP approximation is expected to accurately describe the more standard, early-coupling scenarios. In these cases, the DM distribution $f_\chi(\bs{v})$ is initially broad (as the DM starts hot), and collisions with baryons are hence well described by the FP operator, which allows to follow the evolution of $f_\chi$ from its initial equilibrium form through decoupling. Once the DM becomes sufficiently cold and its distribution narrow relative to the characteristic change in velocity per scattering, the FP approximation no longer accurately describes the evolution of $f_\chi$. However, the very narrowness of $f_\chi$ guarantees that momentum and heat-exchange rates no longer depend on its exact shape, and become closed expressions of the bulk velocity and effective temperature. By construction, the FP operator implies the same closed-form equations, as it is built to give the exact momentum and heat-exchange rate for a given distribution. Therefore, at late time, the Boltzmann-FP equation is guaranteed to give the correct bulk velocity and temperature evolution\footnote{There is a small caveat here: the momentum equation also depends on the gradient of anisotropic stress, which is not a closed function of bulk velocity and temperature. It is likely that this term is always small, as it is suppressed when the DM is tightly coupled, but we defer its detailed study to future work.} and as a consequence, the correct momentum and heat-exchange rates, despite providing an inaccurate detailed form for $f_\chi(\bs{v})$. 

Let us now consider the late-coupling scenario. The DM distribution starts off cold, effectively with zero temperature. As long as it is sufficiently narrow, its temperature and bulk velocities are correctly evolved by the Boltzmann-FP equation -- again, their evolution equations are identical to those obtained in the MB approximation. The underlying distribution $f_\chi$, however, is neither a MB distribution nor is well described by the solution of the Boltzmann-FP equation. Once the DM starts recoupling and its distribution broadens, momentum and heat-exchange rates become dependent on the exact shape of $f_\chi(\bs{v})$. While the FP approximation now becomes accurate, it nevertheless starts with inaccurate ``initial" conditions for $f_\chi(\bs{v})$ at the onset of re-coupling, and the subsequent $f_\chi$ will suffer from this initial inaccuracy, at least while the DM is marginally coupled. 

To rigorously quantify the inaccuracy of the Boltzmann-FP equation for late-coupling, weakly self-interacting DM models, there seems to be no choice but to solve the exact Boltzmann equation, with the full integral collision operator. In the meantime, one should keep in mind that cosmological bounds on such models are uncertain at the factor-of-a-few level. 

\subsection{Self-interactions}

Eventually, for full generality, one should simultaneously account for DM scattering with baryons and self-scattering. Here we briefly discuss the corresponding collision operator, and approximation strategies. 

Inserting Eq.~\eqref{eq:Gamma} into Eq.~\eqref{eq:C_pchi}, with $s = \chi$ we see that the self-interaction collision operator is the following quadratic non-local operator:
\barr
C_{\chi \chi}[f_\chi](\bs{v}) &=& n_\chi \int d^3 v' ~d^3u ~d^3 u'  \mathcal{M}_{\chi \chi}(\bs{v}, \bs{u}; \bs{v}', \bs{u}') \nonumber\\
&& ~~~~\times \left[ f_\chi(\bs{v}') f_\chi(\bs{u}') - f_\chi(\bs{v}) f_\chi(\bs{u})\right], ~~\label{eq:C_chi_chi}
\earr
where $\mathcal{M}_{\chi \chi}$ is defined in Eq.~\eqref{eq:M_chis} and is symmetric under exchange of primed and unprimed variables. This operator conserves DM number, as well as the total DM momentum and energy densities.

It is well known since Boltzmann \cite{Boltzmann_1872} that such a collision operator increases the entropy functional 
\beq
S[f_\chi] \equiv - \int d^3 v f_\chi(\bs{v}) \ln(f_\chi(\bs{v})).
\eeq
We further demonstrate in Appendix \ref{app:entropy} that this is the \emph{only} functional (up to additive and multiplicative factors) of $f_\chi$ that is increased by the collision operator \eqref{eq:C_chi_chi}.

For a given total momentum and energy, this functional is maximized when $f_\chi$ is the MB distribution. At the order-of-magnitude level, we expect this equilibrium distribution to be reached when $n_\chi  \langle \sigma_{\chi \chi} v \rangle \gg H$. In this limit, the evolution of the DM distribution amounts to solving for its mean velocity $\bs{V}_\chi$ and temperature $T_\chi$. If self-interactions are negligible or marginal, however, its full distribution function has to be solved for.  

It would be useful to also have a diffusion approximation for the full collision operator \eqref{eq:C_chi_chi}. The simplest approximation would be of the form \eqref{eq:FP}, with $(\bs{V}_b, T_b) \rightarrow (\bs{V}_\chi, T_\chi)$ -- through these quantities, the FP operator would therefore be implicitly non-linear in $f_\chi$. In order for this diffusion operator to conserve total momentum and energy for any distribution $f_\chi$, it is relatively straightforward to see that the diffusion coefficient $D^{ij}$ must be isotropic and velocity-independent, so that the FP operator takes the form
\beq
C_{\chi \chi}^{\rm FP}[f_\chi] = \frac{D}2 \frac{\partial}{\partial v^i} \left( \frac{\partial f_\chi}{\partial v^i} + \frac{m_\chi}{T_\chi} (v - V_\chi)_i f_\chi \right). \label{eq:C-chi-chi-FP}
\eeq
Computing the rate of change of entropy with this operator, we find, after integration by parts, 
\beq
\frac{d S}{dt} = \frac32 D \frac{m_\chi}{T_\chi} \left(\frac{T_\chi}{3 m_\chi} \int d^3 v f_\chi (\bs{v}) \left(\frac{\partial_{\bs{v}} f_\chi}{f_\chi} \right)^2 - 1\right).\label{eq:dSdt-FP}
\eeq
For two vector function $\bs{X}(\bs{v})$ and $\bs{Y}(\bs{v})$, we define the scalar product 
\beq
\left \langle \bs{X} , \bs{Y} \right \rangle \equiv \int d^3 v f_\chi(\bs{v}) \bs{X}(\bs{v}) \cdot \bs{Y}(\bs{v}), 
\eeq
and the corresponding norm $|| \bs{X}||^2 \equiv \langle \bs{X}, \bs{X} \rangle$. We may rewrite Eq.~\eqref{eq:dSdt-FP} as
\beq
\frac{d S}{dt} = \frac32 D \frac{m_\chi}{T_\chi} \left( \Big{|}\Big{|} \frac{\bs{v}}{3} \Big{|}\Big{|}^2  \Big{|}\Big{|}\frac{\partial_{\bs{v}} f_\chi}{f_\chi}\Big{|}\Big{|}^2 - 1 \right).
\eeq 
From the Cauchy-Schwarz inequality, we find that 
\beq
\frac{d S}{dt} \geq \frac32 D \frac{m_\chi}{T_\chi} \left( \left \langle \frac{\bs{v}}{3}, \frac{\partial_{\bs{v}} f_\chi}{f_\chi} \right \rangle^2  - 1 \right) = 0,
\eeq
since $\left \langle \frac{\bs{v}}{3}, \frac{\partial_{\bs{v}} f_\chi}{f_\chi} \right \rangle = -1$, as follows from integration by parts.

Therefore, the FP operator \eqref{eq:C-chi-chi-FP} not only conserves number, total momentum and energy, it also leads to an increasing entropy functional, for any distribution function $f_\chi$. The last step will be to determine the constant $D$, which, dimensionally, is of order $D \sim n_\chi \langle \sigma_{\chi \chi} v \rangle T_\chi/m_\chi$. We defer to future works a more detailed study of the FP-approximation to self-interactions, and its implementation simultaneously with DM-baryon scattering. 

\section{Conclusions} \label{sec:conclusion}

We have developed a new theoretical formalism to describe the effect of DM scattering with SM particles on linear-cosmology observables.  We have replaced the ideal thermal fluid approximation, standard in this context, by the Boltzmann equation for the DM phase-space density. This allows to extend the range of DM models accurately modeled, to include weakly self-interacting models, whose velocity distribution is non-thermal. 

We have derived a Fokker-Planck (FP) approximation to the collision operator, enforcing that it recovers the exact momentum and heat-exchange rates for any given distribution. This constitutes, to our knowledge, the first derivation of the FP operator for collisions with non-relativistic SM particles. This approximation is more amenable to efficient numerical implementation than the full integral collision operator. We have discussed the limitations of this approximation, as well as interesting extensions, which we shall pursue in future works.

In addition to developing this new formalism, we have presented numerical solutions of the background Boltzmann-FP equation, in the limit of negligible DM self-interactions. We found that the DM velocity distribution can develop order-unity distortions from the usually assumed Maxwell-Boltzmann (MB) distribution. More importantly, we established that the background heat-exchange rate is over-estimated in the MB approximation by factors as large as $\sim 2-3$, especially for light DM particles, and for DM-baryon cross sections with a steep velocity dependence. We will explore the impact of these changes on upper bounds to the DM-baryon cross section from CMB spectral distortions \citep{ACK15} in an upcoming publication. We expect deviations at least as large for the momentum-exchange rate, relevant for CMB-anisotropy and large-scale-structure studies. We derived the perturbed Boltzmann-FP hierarchy that needs to be solved to estimate this rate, and defer its numerical implementation to future work. 

This work opens up a new window in the study of DM properties: instead of implicitly assuming that DM strongly self-interacts, we consider DM-baryon interactions and self-interactions as independent phenomenological parameters. This will broaden the suite of DM models that can be accurately modeled, and tested. Such a detailed and accurate theoretical modeling of DM-baryon interactions is crucial in order to take full advantage of the sensitivity of \emph{Planck} and its successors, and this article takes the first steps in this direction.

\subsection*{Acknowledgements}

I thank Kimberly Boddy, Jens Chluba, Sergei Dubovsky, Vera Gluscevic, Daniel Grin, Chiara Mingarelli, Juli\'an Mu\~noz, Mohammad Safarzadeh, Tristan Smith, Scott Tremaine and Neal Weiner for useful discussions and comments.

\appendix

\section{Power-law cross section for scattering with Helium} \label{app:Helium}

The effect of scattering with a composite nucleus like Helium is encoded by a form factor in the cross section. This form factor was computed in Ref.~\cite{Catena_15} within the effective theory of DM-nucleon interactions. For Helium, it amounts to an exponential suppression of an otherwise power-law cross section, by a factor $\exp\left[- (1 - \hat{n} \cdot \hat{n}') v^2 (\mu ~ a_{\rm He})^2\right]$, where $v$ is the relative velocity, $\mu \equiv m_{\rm He} m_\chi/(m_{\rm He} + m_\chi)$ is the reduced mass of the DM-Helium system, and $a_{\rm He} \approx 1.5$ fm $\approx 7.6$ GeV$^{-1}$. The characteristic relative velocity is such that $v^2 \sim T_b/m_{\rm He} + T_\chi/m_\chi \leq T_b/\mu$, since $T_\chi \leq T_b$, hence
\beq
v^2 (\mu ~ a_{\rm He})^2 \lesssim a_{\rm He}^2 \mu T_b \leq  a_{\rm He}^2 m_{\rm He} T_b \approx \frac{T_b}{4 ~ \textrm{MeV}}.
\eeq
Therefore, non-pointlike effects in Helium-DM scattering are relevant only at temperature greater than a few MeV. In practice, CMB anisotropies and spectral distortions are not sensitive to baryon-DM interactions beyond redshift of a few million, corresponding to keV temperatures. For the problem of interest, the form factor can therefore safely be set to unity, and the cross section with Helium assumed to be a power-law.

\onecolumngrid

\section{Existence and unicity of entropy functional for self-scattering} \label{app:entropy}

It is well-known since Boltzmann \cite{Boltzmann_1872} that collisions between particles in a gas increase the entropy functional \cite{Tolman_38, TerHarr_55}. In this appendix we further show that, up to a multiplicative constant, this is in fact the \emph{only} increasing functional (or $H$-functional) of the form 
\beq
S[f] = \int d^3 v ~\mathcal{S}(f(\bs{v})), \label{eq:S[f]}
\eeq
where $\mathcal{S}$ is a continuous and differentiable function. Note that this is a restricted class of functionals: one could also consider those of the form $\int d^3 v_1d^3 v_2 \mathcal{S}(f(\bs{v}_1), f(\bs{v}_2))$, etc... In particular, functionals of the form \eqref{eq:S[f]} do not include the $K$-functional suggested by Kac \cite{Kac_56}.

The collision operator arising from particle-particle scattering is a non-linear and non-local operator of the form (in the non-relativistic limit):
\barr
C[f](\bs{v}) = \int d^3 v' ~d^3 w~ d^3 w' ~\delta_{\rm D}(\bs{w} + \bs{w}' - \bs{v}  - \bs{v}')  \delta_{\rm D}(w^2 + w'^2 - v^2  - v'^2 ) \Gamma(\bs{w}, \bs{w}' ; \bs{v}, \bs{v}')\left[  f(\bs{w}) f(\bs{w}') -  f(\bs{v}) f(\bs{v}')  \right].~~~~~~\label{eq:chi-chi}
\earr
Here detailed balance (or time-reversal symmetry) implies that the collision rate is symmetric under exchange of the first and last pairs of variables:
\beq
\Gamma(\bs{w}, \bs{w}' ; \bs{v}, \bs{v}') = \Gamma(\bs{v}, \bs{v}' ; \bs{w}, \bs{w}').
\eeq
It is also symmetric under exchanges $\bs{w}' \leftrightarrow \bs{w}$ and $\bs{v}' \leftrightarrow \bs{v}$. 
This collision operator conserves not only the total number of particles, but also their total momentum and energy. 

Let us now compute the time derivative of Eq.~\eqref{eq:S[f]} under the action of collisions:
\barr
\frac{dS}{dt} &=& \int \mathcal{D} \gamma ~ \left[  f(\bs{w}) f(\bs{w}') -  f(\bs{v}) f(\bs{v}')  \right] \dot{\mathcal{S}}(f(\bs{v})),\\
\mathcal{D}\gamma &\equiv& d^3 v ~d^3 v' ~d^3 w~ d^3 w' ~\delta_{\rm D}(\bs{w} + \bs{w}' - \bs{v}  - \bs{v}')  \delta_{\rm D}(w^2 + w'^2 - v^2  - v'^2 ) \Gamma(\bs{w}, \bs{w}' ; \bs{v}, \bs{v}')
\earr
where $\dot{\mathcal{S}} \equiv d \mathcal{S}/df$. Using the symmetries of the integrand, we rewrite this as
\beq
\frac{dS}{dt} = \frac14 \int \mathcal{D} \gamma ~ \left[  f(\bs{w}) f(\bs{w}') -  f(\bs{v}) f(\bs{v}')  \right] \left[\dot{\mathcal{S}}(f(\bs{v})) + \dot{\mathcal{S}}(f(\bs{v}')) - \dot{\mathcal{S}}(f(\bs{w})) - \dot{\mathcal{S}}(f(\bs{w}')) \right].
\eeq 
We now seek functions $\mathcal{S}$ such that the functional $S$ is an increasing function of time for \emph{any} function $f$. This implies that it must satisfy the inequality
\beq
[f_3 f_4 - f_1 f_2] [\dot{\mathcal{S}}(f_1)+ \dot{\mathcal{S}}(f_2) - \dot{\mathcal{S}}(f_3) - \dot{\mathcal{S}}(f_4)] \geq 0, \ \ \forall~ (f_1, f_2, f_3, f_4) > 0. \label{eq:S-cond}
\eeq
This in turn implies that
\barr
\dot{\mathcal{S}}(f_1)+ \dot{\mathcal{S}}(f_2)  \geq \dot{\mathcal{S}}(f_3) + \dot{\mathcal{S}}(f_4) ~~~~~~\forall~ (f_1, f_2, f_3, f_4) \ \ \textrm{such that} \  f_1 f_2 < f_3 f_4, \\
\dot{\mathcal{S}}(f_1)+ \dot{\mathcal{S}}(f_2)  \leq \dot{\mathcal{S}}(f_3) + \dot{\mathcal{S}}(f_4) ~~~~~~\forall~ (f_1, f_2, f_3, f_4) \ \ \textrm{such that} \  f_1 f_2 > f_3 f_4. 
\earr
If the function $\dot{\mathcal{S}}$ is continuous, it must be that
\beq
\dot{\mathcal{S}}(f_1)+ \dot{\mathcal{S}}(f_2) =  \dot{\mathcal{S}}(f_3) + \dot{\mathcal{S}}(f_4)  ~~~~~~\forall~ (f_1, f_2, f_3, f_4) \ \ \textrm{such that} \  f_1 f_2 = f_3 f_4.
\eeq
In particular, setting one of the $f_i$'s to unity we find 
\beq
\dot{\mathcal{S}}(x y) = \dot{\mathcal{S}}(x) + \dot{\mathcal{S}}(y) -\dot{\mathcal{S}}(1),  \ \ \ \forall ~(x, y) > 0.
\eeq
Differentiating with respect to $y$, and then setting $y = 1$, we obtain
\beq
\ddot{\mathcal{S}}(x) = \frac{\textrm{const}}{x},  \ \ \ \forall x > 0,
\eeq
which, upon two integrations, gives us
\beq
\mathcal{S}(f) = - \alpha  f \ln(f) + \beta f + \textrm{const},
\eeq
where $\alpha$ and $\beta$ are arbitrary constants. In order for Eq.~\eqref{eq:S-cond} to be satisfied, one must have $\alpha > 0$. 

To conclude, we have found that the entropy functional
\beq
S[f] \equiv - \int d^3 v f (\bs{v}) \ln(f(\bs{v}))
\eeq
is increased by self-collisions. Furthermore, we have shown that this is the \emph{only} functional of the form \eqref{eq:S[f]} (up to multiplicative and additive constants) which satisfies this condition. 

\twocolumngrid

\bibliography{dm_scat.bib}

\end{document}